\documentclass[english]{IEEEtran}
\usepackage[T1]{fontenc}
\usepackage[latin9]{inputenc}
\usepackage{array}
\usepackage{longtable}
\usepackage{amsmath}
\usepackage{amsthm}
\usepackage{amssymb}
\usepackage{stackrel}
\usepackage{graphicx}
\usepackage{setspace}
\usepackage{pdfpages}
\usepackage{mathtools, cuted}
\usepackage{lipsum, color}
\PassOptionsToPackage{normalem}{ulem}
\usepackage{ulem}

\makeatletter

\providecommand{\tabularnewline}{\\}

\numberwithin{equation}{section}
\numberwithin{figure}{section}
\theoremstyle{plain}
\newtheorem{thm}{\protect\theoremname}
\theoremstyle{definition}
\newtheorem{defn}[thm]{\protect\definitionname}
\theoremstyle{plain}
\newtheorem{lem}[thm]{\protect\lemmaname}
\theoremstyle{plain}
\newtheorem{cor}[thm]{\protect\corollaryname}
\theoremstyle{definition}
\newtheorem{example}[thm]{\protect\examplename}
\theoremstyle{remark}
\newtheorem{claim}[thm]{\protect\claimname}

\usepackage{tikz}
\usetikzlibrary{positioning}
\usepackage{pgfplots}
\newdimen\R 
\newdimen\SmallR 
\makeatother

\usepackage{babel}
\providecommand{\claimname}{Claim}
\providecommand{\corollaryname}{Corollary}
\providecommand{\definitionname}{Definition}
\providecommand{\examplename}{Example}
\providecommand{\lemmaname}{Lemma}
\providecommand{\theoremname}{Theorem}

\begin{document}
\title{Sub-Rate Linear Network Coding}

\author{B. Grinboim, I. Shrem and O. Amrani}
\maketitle
\begin{abstract}
\begin{singlespace}
Increasing network utilization is often considered as the holy grail
of communications. In this article, the concept of subrate coding
and decoding in the framework of linear network coding (LNC) is disscussed
for single-source multiple-sinks finite acyclic networks. Sub-rate
coding offers an add-on to existing LNC. It allows sinks whose max-flow
is smaller than the source message-rate, termed \emph{sub-rate sinks},
to decode a portion of the transmitted message without degrading the
maximum achievable rate of LNC sinks whose max-flow is equal (or greater)
than the rate of the source node. The article studies theoretical
aspects of sub-rate coding by formulating the conditions a node (and
indeed the network) must fulfill so as to qualify as a legitimate
sub-rate sink.
\end{singlespace}
\end{abstract}

\section{Introduction}

\begin{singlespace}
Network coding is a research field targeting the improvement of communications
within a network \cite{Farooqi}. The fundamental idea, discussed
in \cite{Alshwede}, is that intermediate nodes in the network serve
as computational units rather than merely relays (i.e. assuming computational
tasks may be implemented in them), thus supporting increased overall
network throughput. Network coding has beenown to potentially increase
overall network throughput in single-source multiple-sink communications
scenarios (i.e., broadcast of massages), or multi-source networks\cite{LiFeb.2003}.
One of the fundamental results in this field is that in the case of
a single-source, acyclic, and finite network, given that all computations
are done over a field large enough, LNC can achieve the maximum capacity
to every \emph{eligible sink} \cite{LiFeb.2003,KoetterOct.2003}.
By eligible sink, we refer to a sink whose max-flow (achievable throughput)
is greater or equal to the source transmission-rate.

A simple example demonstrating the strength of a network coding is
the butterfly network, a network that transmits two signals $a$ and
$b$ to two sinks, $6$ and $7$ as demonstrated in Figures \ref{fig:Butterfly-network-without}
and \ref{fig:Butterfly-network-with}. In Figure \ref{fig:Butterfly-network-without},
the network does not employ network coding (i.e., intermediate nodes
do not execute any coding-related computations) - hence only one of
the sinks, $6$ or $7,$ can decode both messages in each network
use. In Figure \ref{fig:Butterfly-network-with}, the network employs
network coding. To this end, node $4$ can XOR (exclusive-or) both
its inputs, which in turn enables both sinks to decode the two transmitted
messages.

\begin{figure}
\begin{minipage}[c]{0.45\columnwidth}%
\begin{center}
\caption{\label{fig:Butterfly-network-without}Butterfly network without network
coding}
\begin{tikzpicture}[ 
regularnode/.style={circle, draw=black, fill=white, thick, minimum size=7mm}, 
locator/.style={circle, draw=white, fill=white,thin, minimum size=1mm},
node distance=0.5cm
] 
\node[regularnode] (1) {1};
\node[locator] (under1) [below=of 1] {};
\node[regularnode] (2) [right=of under1] {3};
\node[regularnode] (3) [left=of under1]  {2};
\node[locator] (under3) [below=of 3] {};
\node[regularnode] (4) [right=of under3]  {4};
\node[regularnode] (5) [below=of 4]  {5};
\node[locator] (under5) [below=of 5] {};
\node[regularnode] (6) [left=of under5]  {6};
\node[regularnode] (7) [right=of under5]  {7};
\draw[->] (1.south east) -- (2.north west) node[near start,right] {b};
\draw[->] (1.south west) -- (3.north east) node[near start,left] {a}; 
\draw[->] (2.south west) -- (4.north east)
node[near start,left] {b};
\draw[->] (3.south east) -- (4.north west) node[near start,right] {a};
\draw[->] (4.south) -- (5.north) node[midway,right] {a/b}; 
\draw[->] (5.south east) -- (7.north west) node[near start,right] {a/b}; 
\draw[->] (5.south west) -- (6.north east) node[near start,left] {a/b};
\draw[->] (2.south east) -- (7.north east) node[midway,right] {b};
\draw[->] (3.south west) -- (6.north west) node[midway,left] {a};
\end{tikzpicture} 
\par\end{center}%
\end{minipage}\hfill{}%
\begin{minipage}[c]{0.45\columnwidth}%
\begin{center}
\caption{\label{fig:Butterfly-network-with}Butterfly network with network
coding}
\begin{tikzpicture}[ 
regularnode/.style={circle, draw=black, fill=white, thick, minimum size=7mm}, 
locator/.style={circle, draw=white, fill=white,thin, minimum size=1mm},
node distance=0.5cm
] 
\node[regularnode] (1) {1};
\node[locator] (under1) [below=of 1] {};
\node[regularnode] (2) [right=of under1] {3};
\node[regularnode] (3) [left=of under1]  {2};
\node[locator] (under3) [below=of 3] {};
\node[regularnode] (4) [right=of under3]  {4};
\node[regularnode] (5) [below=of 4]  {5};
\node[locator] (under5) [below=of 5] {};
\node[regularnode] (6) [left=of under5]  {6};
\node[regularnode] (7) [right=of under5]  {7};
\draw[->] (1.south east) -- (2.north west) node[near start,right] {b};
\draw[->] (1.south west) -- (3.north east) node[near start,left] {a}; 
\draw[->] (2.south west) -- (4.north east)
node[near start,left] {b};
\draw[->] (3.south east) -- (4.north west) node[near start,right] {a};
\draw[->] (4.south) -- (5.north) node[midway,right] {a+b}; 
\draw[->] (5.south east) -- (7.north west) node[near start,right] {a+b}; 
\draw[->] (5.south west) -- (6.north east) node[near start,left] {a+b};
\draw[->] (2.south east) -- (7.north east) node[midway,right] {b};
\draw[->] (3.south west) -- (6.north west) node[midway,left] {a};
\end{tikzpicture} 
\par\end{center}%
\end{minipage}
\end{figure}

In previous works pertaining to network coding, sinks with smaller
max-flow than the transmission rate, (herein termed sub-rate sinks)
had been considered ineligible, and so communication to them was avoided.
This approach is derived from the classical definition of a multicast:
a sink that can not receive all the information (a.k.a. network rate)
is considered ``useless''. While this holds true for many cases,
there are other cases where retrieving some of the transmitted data
is valuable; for example when a sub-rate sink can benefit from reduced
data ``resolution''. Alternatively, one can look at it the other
way around - the ability to communicate with some network sinks in
the multicast using higher rate than other sinks, may enable the higher-rate
sinks to retrieve the transmitted information faster than the others
(given the communication is not continuous), or use the ``spare''
capacity for conveying data that is not necessarily required by all
the sinks, e.g. monitoring information.

The question therefore: is it possible to configure a network code
such that some of the ineligible sinks can still decode a portion
of the input data? apparently, in some cases - it is certainly possible.
An example is given in Figures \ref{fig:modified butterfly without}
and \ref{fig:modifired butterfly with}. In this example, another
sink, $8,$ is added to the butterfly network, a node whose max-flow
(from the source) is only $1$. Clearly, it is assumed to be beneficial
(hence required) to communicate with node $8$ at rate of $1$ symbol
per network use. In the original configuration of the butterfly network
code - node $8$ receives $a+b$ - and hence cannot decode any symbol,
as shown in Figure \ref{fig:modified butterfly without}. By properly
manipulating the transmitted symbols at the source node though, as
depicted in Figure \ref{fig:modifired butterfly with}, it is possible
to configure the network so that node $8$ operates at a communication
rate of $1$ symbol per network use, without affecting the ability
of eligible sinks, $6$ and $7$, to operate at their original rate
of $2$ symbols per network use.

\begin{figure}
\centering{}%
\begin{minipage}[c]{0.45\columnwidth}%
\begin{center}
\caption{\label{fig:modified butterfly without}Modified butterfly network
without sub-rate coding}
\begin{tikzpicture}[ 
regularnode/.style={circle, draw=black, fill=white, thick, minimum size=7mm}, 
locator/.style={circle, draw=white, fill=white,thin, minimum size=1mm},
node distance=0.5cm
] 
\node[regularnode] (1) {1};
\node[locator] (under1) [below=of 1] {};
\node[regularnode] (2) [right=of under1] {3};
\node[regularnode] (3) [left=of under1]  {2};
\node[locator] (under3) [below=of 3] {};
\node[regularnode] (4) [right=of under3]  {4};
\node[regularnode] (5) [below=of 4]  {5};
\node[locator] (under5) [below=of 5] {};
\node[regularnode] (6) [left=of under5]  {6};
\node[regularnode] (7) [right=of under5]  {7};
\node[regularnode] (8) [below=of under5]  {7};
\draw[->] (1.south east) -- (2.north west) node[near start,right] {b};
\draw[->] (1.south west) -- (3.north east) node[near start,left] {a}; 
\draw[->] (2.south west) -- (4.north east)
node[near start,left] {b};
\draw[->] (3.south east) -- (4.north west) node[near start,right] {a};
\draw[->] (4.south) -- (5.north) node[midway,right] {a+b}; 
\draw[->] (5.south east) -- (7.north west) node[near start,right] {a+b}; 
\draw[->] (5.south west) -- (6.north east) node[near start,left] {a+b};
\draw[->] (2.south east) -- (7.north east) node[midway,right] {b};
\draw[->] (3.south west) -- (6.north west) node[midway,left] {a};
\draw[->] (5.south) -- (8.north) node[midway,left] {a+b};
\end{tikzpicture} 
\par\end{center}%
\end{minipage}\hfill{}%
\begin{minipage}[c]{0.45\columnwidth}%
\begin{center}
\caption{\label{fig:modifired butterfly with}Modified butterfly network with
sub-rate coding}
\begin{tikzpicture}[ 
regularnode/.style={circle, draw=black, fill=white, thick, minimum size=7mm}, 
locator/.style={circle, draw=white, fill=white,thin, minimum size=1mm},
node distance=0.5cm
] 
\node[regularnode] (1) {1};
\node[locator] (under1) [below=of 1] {};
\node[regularnode] (2) [right=of under1] {3};
\node[regularnode] (3) [left=of under1]  {2};
\node[locator] (under3) [below=of 3] {};
\node[regularnode] (4) [right=of under3]  {4};
\node[regularnode] (5) [below=of 4]  {5};
\node[locator] (under5) [below=of 5] {};
\node[regularnode] (6) [left=of under5]  {6};
\node[regularnode] (7) [right=of under5]  {7};
\node[regularnode] (8) [below=of under5]  {7};
\draw[->] (1.south east) -- (2.north west) node[near start,right] {a+b};
\draw[->] (1.south west) -- (3.north east) node[near start,left] {a}; 
\draw[->] (2.south west) -- (4.north east)
node[near start,left] {a+b};
\draw[->] (3.south east) -- (4.north west) node[near start,right] {a};
\draw[->] (4.south) -- (5.north) node[midway,right] {b}; 
\draw[->] (5.south east) -- (7.north west) node[near start,right] {b}; 
\draw[->] (5.south west) -- (6.north east) node[near start,left] {b};
\draw[->] (2.south east) -- (7.north east) node[midway,right] {a+b};
\draw[->] (3.south west) -- (6.north west) node[midway,left] {a};
\draw[->] (5.south) -- (8.north) node[midway,left] {b};
\end{tikzpicture} 
\par\end{center}%
\end{minipage}
\end{figure}

This work studies the ability to communicate with some of the network
nodes as sub-rate max-flow sinks (without affecting the eligible sinks)
using tailored linear network codes.
\end{singlespace}

The rest of the paper is organized as follows. Section 2 provides
the notations that will be used throughout the paper. Section 3 presents
the basic intuition of sub-rate coding with LNC, which motivates the
following definitions and theoretical contributions. The methodology
of \emph{ful sub-rate coding} is given in Section 4 along with sufficient
conditions for its existence in a linear multicast with a set of sub-rate
designated sinks. An algorithm for realizing such a linear multicast
is also provided. Section 5 generalizes this concept and defines \emph{partial
sub-rate coding}, i.e. resorting to blocks of a few messages in those
cases where full sub-rate coding is not possible. In Section 6, further
discussion on the presented results, as well as their combination
with other LNC classes are discussed. Section 7 concludes the paper.

\section{Notations }
\begin{singlespace}

\subsection{Linear Network Coding}
\end{singlespace}

\begin{singlespace}
Consider a network $G'=\{\mathcal{V},E\}$, with $\mathcal{V}$ and
$E$ denoting network nodes and links (alternative terms for vertices
and edges as we are dealing with a communication network), respectively;
a source node $s$ in $\mathcal{V}$ and a set of $d$ sink nodes
$T=\{t_{1},\ldots t_{d}\}$ in $\mathcal{V}$. The max-flow between
$s$ and $t_{i}\in$$T$ is denoted by $h_{t_{i}}$; in this paper,
the term ``max-flow'' of a sink $t_{i}$ shall always refer to this
$h_{t_{i}}$. All the information symbols are regarded as elements
of a base field $F$ (this terminology is often referred to as \emph{scalar
network codes}). 
\end{singlespace}

The \emph{rate }of the communication network, denoted by $r$, is
a positive integer representing the number of symbols created by the
source every network use. In this paper, assume that $r$ is constant. 

For every node $x\in\mathcal{V}\setminus\{s\}$, denote by $In(x)$
and $Out(x)$ the sets of input and output edges to the node $x$,
respectively. For convenience, it is customary to add another node
$s'$, referred to as the \emph{imaginary source}, that has $r$ outgoing
edges to the original source $s$ - referred to as the $r$ \emph{imaginary
links} (namely, $In(s)$ is the set of $r$ imaginary links). Assume
that $G'$ includes the imaginary source and edges.

The basic concept of network coding is that all the nodes in the network
are able to perform calculations over the field $F$. 

\begin{singlespace}
The following is a description of a \emph{linear network code} (LNC)
for finite acyclic network that is derived from \cite{Yeung2006}:
\end{singlespace}
\begin{defn}
\begin{singlespace}
\label{def:GEK} An $r$-dimensional $F$-valued linear network code
operating in an acyclic communication network is characterized by:
\end{singlespace}
\begin{enumerate}
\item \emph{Local Encoding Kernel }(LEK) - $\{k_{d,e}\}$ - A set of scalars
$k_{d,e}\in F$ for every adjacent pair of edges $(d,e)$ in the network;
\item \emph{Global Encoding Kernel} (GEK) - $\{f_{e}\}_{e\in E}$ - A set
of $r$-dimensional column vectors $f_{e}$ for every edge $e$ in
the network such that:
\begin{enumerate}
\begin{singlespace}
\item For every non-source node $x$, and every $e\in Out(x)$, $f_{e}=\sum_{d\in In(x)}k_{d,e}f_{d}$;
\item For the imaginary links $e\in In(s)$, the set of vectors $\{f_{e}\}_{e\in In(s)}$
are the $r$ linearly independent vectors that constitute the natural
basis of the vector space $F^{r}$.
\end{singlespace}
\end{enumerate}
\end{enumerate}
\begin{singlespace}
The local encoding kernel (LEK) associated with node $x$ refers to
a $|In(x)|\times|Out(x)|$ matrix. The vector $f_{e}$ is called the
\emph{global encoding kernel} (GEK) for edge $e$. 
\end{singlespace}
\end{defn}
\begin{singlespace}
Note that given the local encoding kernels at all the nodes in an
acyclic network, the global encoding kernels can be calculated recursively
in an upstream-to-downstream order according to the given definition. 
\end{singlespace}
\begin{defn}
\begin{singlespace}
\label{def:linear multicastt}Let $\{f_{e}\}_{e\in E}$ denote the
global encoding kernels in an $r$-dimensional $F$-valued linear
network code on a single-source finite acyclic network. Let $V_{x}=span\{f_{d}|d\in In(x)\}$.
Then, the linear network code is a \emph{linear multicast} if $dim(V_{x})=r$
for every non-source node $x$ with $max-flow(x)\geq r.$
\end{singlespace}
\end{defn}
\begin{singlespace}
In \cite{LiFeb.2003,Yeung2006}, an algorithm for the construction
of a linear multicast is provided, requiring the field size $|F|$
to be greater than the size of the set of sinks (whose $maxflow(\{T\})\geq r$)
i.e, $|F|>|\{T:maxflow(\{T\})\geq r\}|$. This algorithm provides
a LNC that allows all target sinks to receive decodable information
simultaneously in every network use.

Note that this algorithm does not require that the set of target sinks
include \emph{all} the \emph{eligible sinks} (satisfying $h_{t_{i}}\geq r$).
i.e., it is also possible to use the algorithm to create a \emph{generalized
linear multicast}, which is defined as follows:
\end{singlespace}
\begin{defn}
\begin{singlespace}
\label{def:generalized linear multicast}Let $\{f_{e}\}_{e\in E}$
denote the global encoding kernels in an $r$-dimensional $F$-valued
linear network code on a single-source finite acyclic network. Let
$V_{x}=span\{f_{d}|d\in In(x)\}$. Define a set of non-source nodes
$T\subseteq\mathcal{V}$ as sinks. Then, the linear network code is
a \emph{generalized linear multicast} if $dim(V_{t})=r$ for any sink
$t\in T$ with $max-flow(t)\geq r.$
\end{singlespace}
\end{defn}
\begin{singlespace}
For convenience, in this paper any reference to \emph{linear multicast
}refers to the generalized definition.

Note that not all the links in $G'$ need necessarily be employed
by the network code in the case of a generalized linear multicast.
Hereinafter, we shall only be referring to the network $G$, which
is defined as the sub-network of $G'$ consisting only of the links
from $E$ that are employed by the network code along with their adjacent
nodes. A network node in $G$ that is neither source, nor sink, shall
be called an \emph{intermediate node}. A \emph{path} is a set of links
that provides a connection between two nodes in the network.
\end{singlespace}

As defined in \ref{def:generalized linear multicast}, a linear multicast
assumes that every network use the source is generating and transmitting
$r$ symbols, and every eligible sink $t_{i}$ can extract the transmitted
symbols from the $h_{t_{i}}$different symbols it receives. Nevertheless,
in this paper, a node $t_{i}$ could be defined as a sink of the multicast
even if $h_{t_{i}}<r$. In this case, the algorithm described in \cite{LiFeb.2003,Yeung2006}
for constructing a linear multicast yields $h_{t_{i}}$ independent
paths from $s$ to $t_{i}$, resulting in $h_{t_{i}}$ linearly independent
symbols received by $t_{i}$ every network use.

\begin{singlespace}
Two different approaches to LNC had been proposed: modeling the data
units and the coding operations over the finite field $GF(q^{L})$
(scalar network coding), or modeling the data units and vector operations
over $GF(q)^{L}$ (vector network coding) \cite{Ebrahimi2010.June,KoetterOct.2003}.
For simplicity, in this manuscript we are focused with scalar network
coding, but all the results may be equivalently applicable to vector
network coding.
\end{singlespace}

\subsection{Global Encoding Kernel and Matrix}

For the standard scenario whereby a network operates at a constant
rate $r$, a Global Encoding Kernel (GEK) - defined as a per-node
function that determines the information conveyed by each network
link as a function of the network input - can be derived from the
LEK (and the other way around). Consequently, a \emph{linear }network
code is equivalently defined by a Global Encoding Matrix (GEM) $B,$
a $r\times m$ matrix over the field $F$, where $m=|E|$. Every column,
$b_{e}$, of $B$ is associated with one network link, such that multiplying
an input vector $v\in F^{r}$ by this column results in the symbol
from $F$ that the network code conveys on that link per network input
vector $v$. Observing that the $h_{t}$ links impinging on the sink
$t$ are employed for decoding the information reaching this sink,
one can denote by $B_{t}$ the $r\times h_{t}$ matrix whose columns
are given by $\{b_{e}|e\ is\ entering\ t\ \&\ used\ for\ decoding\:in\ t\}$.
$B_{t}$ will be called the global encoding matrix - \emph{GEM} -
of the sink $t$. Again, from the linear multicast definition, it
is guaranteed that the columns of $B_{t}$ are linearly independent.
Particularly, if a sink $t$ satisfies $h_{t}\geq r$, $B_{t}$ is
an invertible matrix (the sink only uses $min\{h_{t},r\}$ links for
decoding). For convenience, the vector space spanned by the columns
of $B_{t}$ will be denoted by $B_{t}'$.

\subsection{Decoding}

Each sink $t$ holds a $h_{t}\times h_{t}$ decoding matrix, $D_{t}$,
whose entries are defined over the field $F$. The decoding matrix
is used for extracting the source information $v$ from the sink input,
$vB_{t}$. When $h_{t}\geq r$, $B_{t}$ is invertible, and hence,
the natural choice of $D_{t}$ is given by $D_{t}=B_{t}^{-1}$; thus
clearly $vB_{t}D_{t}=v$. The case $h_{t}<r$ is generally not decodable,
and was not considered before (to the best of our knowledge). This
provides the motivation for the following.

\subsection{Precoding}

Next, we wish to introduce precoding into the framework of network
coding. Precoding is to be executed at the source node, prior to transmission.
To this end, the source information $v$ shall be multiplied by an
$r\times r$ matrix $P$ over the field $F,$hence termed \emph{precoding},
such that the input to the network shall be $vP$ rather than $v$.
For the data to be decodable by the sinks, $P$ must be invertible
(i.e., the $r$ outputs from the source must be linearly independent).
Consequently, for each sink $h_{t}$ with $h_{t}\geq r$, the choice
of decoding matrix $D_{t}$ will change to $D_{t}=B_{t}^{-1}P^{-1}$,
thus providing $vPB_{t}D_{t}=v$. Given that, the network code and
the precoding scheme are independent with respect to one another.
Meaning that the construction of the network code is carried out irrespective
of the precoding, the precoding process is reversible, so its effect
on the information can clearly be mitigated by those sinks satisfying
$h_{t}\geq r$.

\subsection{\label{subsec:Sub-rate-Block-LNC}Sub-Rate Block LNC}

Finally, $l$ consecutive network uses may be referred to as a block
LNC of length $l\cdot r$. We relate to this approach as the $l$-block
linear multicast of $B$. In this paper, the network code and the
transmission rate are considered constant within a block. Therefore:
\begin{enumerate}
\item A row vector $\hat{v}$ of length $l\cdot r$ is the, so-called, \emph{LNC
message block} at the network input; 
\item The precoding matrix $\hat{P}$ is an invertible $l\cdot r\times l\cdot r$
matrix;
\item The global encoding kernel matrix is a $l\cdot r\times l\cdot|E|$
block matrix $\hat{B}$, with $l$ identical blocks $B$ on the diagonal.
Accordingly, each sink $t$ is reached via a global encoding matrix
$\hat{B_{t}}$ which is a $l\cdot r\times l\cdot h_{t}$ block matrix;
\item The decoding matrix of the sink $t$, $\hat{D_{t}}$, is a $l\cdot h_{t}\times l\cdot h_{t}$
matrix. The decoding rate of a sink is defined as $r_{t}=\#\{decodable\ symbols\}/l$.
It is fairly easy to see that for a sink $t$ with $h_{t}\geq r$
(i.e $B_{t}$ is invertible), $\hat{B_{t}}$ is also invertible. Therefore,
defining $\hat{D_{t}}=\hat{B_{t}}^{-1}\hat{P}^{-1}$ allows $t$ to
decode all the transmitted data, thus supporting a decoding rate of
$\frac{l\cdot r}{l}=r$, just as in the single-network-use case.
\end{enumerate}

\section{Motivation and Basic Approach}

Sub-rate coding and decoding is introduced for improving the overall
system throughput achievable via linear multicast. Such improvement
can be realized if proper coding and decoding approach is found so
as to allow those nodes whose max-flow, $h_{t}$, is smaller than
the source rate $r$, to extract at least a portion of $r$ data units
in each network use.

In the sequel, a sink $t$ with $h_{t}<r$, will be referred to as
\emph{sub-rate sink}. According to information theory, a sink cannot
operate at rates higher than its max-flow $h_{t}$ (the max-flow between
the source and a sink is essentially the capacity of the corresponding
set of links constituting a communication path). It is, therefore,
the aim of the proposed approach to provide communication rate as
close as possible to the max-flow of a sub-rate sink, while allowing
no interference, or rate loss, to be experienced by any eligible sink.
Note that the max-flow to a sub-rate sink $t$ is sink dependent,
and different sub-rate sinks may have different max-flows.

In this section, a few basic scenarios of sub-rate coding shall be
explicitly demonstrated in order to shed light on the proposed coding
methodology. All the examples concern a specific case of a network
operating at symbol generation rate of $r=3$.

\subsection{A Single Sub-Rate Sink }

To begin with, the case of a single sub-rate sink $t$ (with $h_{t}<r$)
shall be considered. In this case, the GEM of $t$, $B_{t}$, is a
$3\times r$ full (column) rank matrix.
\begin{lem}
Let sink $t$ be a single sub-rate sink, with $h_{t}=2$ in a linear
multicast with $r=3$. There exists a linear multicast of rate $3$
allowing $t$ to work at decoding rate of $2$.
\end{lem}
Before formally proving the lemma, the following definition is inserted:
\begin{defn}
\label{def:a-BR-factorization-of}a \emph{BR-factorization} of an
arbitrary $m\times n$ matrix $A$, ($m\geq n$), with linearly independent
columns, is defined as a multiplication of two matrices $A=BR$, where
$B$ is an invertible $m\times m$ matrix, and $R$ is an $m\times n$
matrix whose columns are \emph{n} out of \emph{m} (distinct) columns
taken from the identity $m\times m$ matrix $I_{m}$.
\end{defn}
It is easy to see that any full rank matrix has a BR-factorization:
consider e.g. $B=(A\ |\ A^{c});A^{c}$ being an $m\times(m-n$) complimentary
matrix such that $B$ spans the \emph{m}-dimensional space over the
field \emph{F}, and $R=(The\ first\ n\ columns\ of\ I_{m})$.

Note that BR-factorization of a matrix $A$ is not unique.
\begin{IEEEproof}
Using a BR-factorization of $B_{t}$, we can define $B_{t}=\bar{B}R$.
By definition, $\bar{B}$ is invertible. Taking $P=\bar{B}^{-1}$
provides:
\begin{enumerate}
\item Any sink $t'$ with $h_{t'}\geq3$, can fully decode the transmitted
data, with the choice of decoding matrix $D_{t'}=B_{t'}^{-1}P^{-1}$,
since $PB_{t'}D_{t'}=I_{3}$.
\item The sub-rate sink $t$ can operate with decoding rate of $h_{t}=2$,
with the choice of $D_{t}=I_{2}$, $PB_{t}D_{t}=\bar{B}^{-1}\bar{B}RI_{2}=R$.
Recall that $R$ is a $3\times2$ matrix with columns from $I_{3}$
-- and so $t$ can decode two of the three transmitted symbols.
\end{enumerate}
\end{IEEEproof}
The case of $h_{t}=1$ is simple, and can be treated in exactly the
same way. Obviously, the decoding rate of $t$ can only be $1$ in
this case.

\subsection{Two Sub-Rate Sinks}
\begin{lem}
In the case of two sub-rate sinks $t_{1}$ and $t_{2}$, there exists
a linear multicast of rate $3$ allowing $t_{1}$ and $t_{2}$ to
work at decoding rates of $h_{t_{1}}$ and $h_{t_{2}}$ respectively.
\end{lem}
\begin{IEEEproof}
The proof will be given by examining all the possible cases:
\begin{enumerate}
\item \textbf{$\boldsymbol{h_{t_{1}}=h_{\boldsymbol{t}_{\boldsymbol{2}}}=1}$.}
It is shown that there exists BR-factorizations $B_{t_{i}}=\bar{B}R_{i},\ i=1,2$,
that satisfy \\ $\bar{B_{1}}=\bar{B_{2}}=\bar{B}$:
\begin{enumerate}
\item If the vectors $B_{t_{1}},B_{t_{2}}$ are linearly dependent, $B_{t_{1}}=a\cdot B_{t_{2}}$,
$\bar{B}$$=(B_{t_{1}}|two \ CLIV),\ P=\bar{B}^{-1}$,
with $D_{t_{1}}=1$ and $D_{t_{2}}=a$, provides $PB_{t_{i}}D_{t_{i}}=R_{t_{i}}$ (with CLIV = {\it complimentary linearly independent vectors}).
In this case, both sub-rate sinks can decode the same symbol.
\item If the vectors $B_{t_{1}},B_{t_{2}}$ are linearly independent, the
choice $\bar{B}=(B_{t_{1}}|B_{t_{2}}|CLIV)$,
$D_{t_{1}}=D_{t_{2}}=1$ provides $PB_{t_{i}}D_{t_{i}}=R_{t_{i}}$.
In this case, each sub-rate sink decodes a different symbol.
\end{enumerate}
\item $\boldsymbol{h_{t_{i}}=1,h_{t_{j}}=2}$ (w.l.o.g, $i=1,j=2$):
\begin{enumerate}
\item If the vector $B_{t_{1}}$ is linearly independent of (i.e. not on
the ``plane'' spanned by) the vectors of $B_{t_{2}}$, then choosing
$\bar{B}=(B_{t_{1}}|two\ vectors\ spanning\ B_{t_{2}}),\ P=\bar{B}^{-1},\,D_{t_{1}}=1,\:D_{t_{2}}=I_{2}$
shall provide $PB_{t_{i}}D_{t_{i}}=R_{t_{i}}$.
\item If the vector $B_{t_{1}}$ is on the plane spanned by the vectors
of $B_{t_{2}}$, a slightly different approach is used: essentially,
decoding matrix $D_{t_{2}}$   is required to manipulate the columns
of $B_{t_{2}}$ so that one of them becomes $B_{t_{1}}$, but the
span of the obtained matrix does not change. Stated mathematically,
$span\{B_{t_{2}}D_{t_{2}}\}=span\{B_{t_{2}}\}$, while $B_{t_{1}}$
is one of the columns of $B_{t_{2}}D_{t_{2}}$. Such $D_{t_{2}}$
necessarily exists since $B_{t_{1}}$ is in $span\{B_{t_{2}}\}$.
Then, the choice of $\bar{B}=(B_{t_{2}}D_{t_{2}}|CLIV),\ P=\bar{B}^{-1},D_{t_{1}}=1$
provide $PB_{t_{i}}D_{t_{i}}=R_{t_{i}}$. Notably, $B_{t_{2}}=\bar{B}R_{t_{2}}D_{t_{2}}^{-1}$.
\end{enumerate}
\item \textbf{$\boldsymbol{h_{t_{1}}=h_{t_{2}}=2}$}:
\begin{enumerate}
\item In case the vectors in $B_{t_{1}}$ and $B_{t_{2}}$ span the same
subspace, one can pick any two linearly independent vectors $v_{1}$
and $v_{2}$ in that subspace, then choose $\bar{B}=(v_{1}|v_{2}|CLIV)$.
Clearly, matrices $D_{t_{1}}$ and $D_{t_{2}}$ exist such that for
$i=1,2$, the columns of $B_{t_{i}}D_{t_{i}}$will be $(v_{1}|v_{2})$.
As in the previous case, $B_{t_{i}}=\bar{B}R_{t_{i}}D_{t_{i}}^{-1}$.
Therefore $P=\bar{B}^{-1}$ alongside the corresponding $D_{t_{i}}$
provide $PB_{t_{i}}D_{t_{i}}=R_{t_{i}}$.
\item In case the vectors in $B_{t_{1}}$ and $B_{t_{2}}$ span different
subspaces, these subspaces define two different planes. Therefore,
their intersection forms a line. Let $v$ be a vector on that line.
Let $v_{1}$ and $v_{2}$ be two vectors in $B_{t_{1}}$ and $B_{t_{2}}$
respectively, such that $v_{i}\in span\{B_{t_{i}}\}$ for $i=1,2$,
and $\{v,v_{1},v_{2}\}$ is a linearly independent set. Choosing $\bar{B}=(v|v_{1}|v_{2})$,
decoding matrices $D_{t_{1}}$ and $D_{t_{2}}$ exist, such that for
$i=1,2$, the columns of $B_{t_{i}}D_{t_{i}}$ will be $(v|v_{i})$.
Therefore $P=\bar{B}^{-1}$ alongside these $D_{t_{i}}$ provide $PB_{t_{i}}D_{t_{i}}=R_{t_{i}}$.
\end{enumerate}
\end{enumerate}
\end{IEEEproof}
Based on the understanding of the methodology gained in the aforementioned
scenarios, the following lemma is direct.
\begin{lem}
\label{lem:LBR_fact_of_BD}Let $B\in M(F)_{r\times m}$ be a global
encoding matrix for a linear multicast on the graph $G=\{\mathcal{V},E\}$
with a source $s\in\mathcal{V}$ and set of sinks $T\subseteq\mathcal{V}$,
where $m=|E|$. For every sink $t\in T,$ denote by $B_{t}$ the GEM
of $t$, by $h_{t}$ its max-flow from the source $s$, and by $D_{t}$
its decoding matrix. Let $T'\subset T$ be a set of sub-rate sinks,
$\{t_{1},\ldots,t_{d}\}$. If there exists a matrix $\bar{B}$, and
matrices $D_{t}$ and $R_{t}$ for each sub-rate sink $t$, such that
$\bar{B}R_{t}$ is a BR factorization of $B_{t}D_{t}$, then each
sub-rate sink $t\in$$T'$ can work at a decoding rate $h_{t}$.
\end{lem}
\begin{IEEEproof}
The matrix $\bar{B}$ is invertible by the definition of the BR-factorization.
Hence, the precoding matrix $P$ can be defined as $P=\bar{B}^{-1}$.
The input to a sink $t$ will then be $PB_{t}$; multiplying this
by the decoding matrix of $t$, $D_{t}$, the output of $t$ is: $PB_{t}D_{t}=\bar{B}^{-1}\bar{B}R_{t}=R_{t}$.
Since the columns of $R_{t}$ are, by the definition of the BR-factorization,
different columns of $I_{r}$, then given a transmitted data vector
$v\in F^{r}$, $vPB_{t}D_{t}=vR_{t}$ - and $t$ is capable of working
at a decoding rate $h_{t}$.
\end{IEEEproof}
\begin{defn}
\label{def:exact_spanner}A set of vectors $V=\{v_{1},...,v_{d}\}$
will be called an \emph{exact spanner} with respect to the set of
matrices $\tilde{B}=\{B_{1},...,B_{k}\}$, if for each and every $r\times h_{i}$
matrix $B_{i}\in\tilde{B}$, there exists a subset of $V$ with exactly
$h_{i}$ vectors $\{v^{1},...,v^{h_{i}}\}\subset V$, with $span\{v^{1},...,v^{h_{i}}\}=span\{B_{i}\}\stackrel{def}{=}B_{i}'$.
Note that an exact spanner $V$ may contain linearly dependent vectors.
\end{defn}
\begin{cor}
\label{cor:exact spanner_usage}Let $B\in M(F)_{r\times m}$ be a
global encoding matrix in a linear multicast on the graph $G=\{\mathcal{V},E\}$
with a source $s\in\mathcal{V}$ and set of sinks $T\subseteq\mathcal{V}$.
For every sink $t\in T$, denote by $B_{t}$ the GEM of $t$, by $h_{t}$
its max-flow from the source $s$, and by $D_{t}$ its decoding matrix.
Let $T'\subset T$ be a set of sub-rate sinks, $\{t_{1},\ldots,t_{d}\}$,
with $\tilde{B}=\{B_{t_{1}},...,B_{t_{d}}\}$. If there exists a set
of $r$ vectors $U=\{v_{1},\text{\dots},v_{r}\}$ such that:
\begin{enumerate}
\item $\{v_{1},\text{\dots},v_{r}\}$ are linearly independent;
\item $U$ is an exact spanner with respect to $\tilde{B}$,
\end{enumerate}
then each $t_{i}$$\in T'$ can work at a decoding rate $h_{t_{i}}$.
\end{cor}
\begin{IEEEproof}
let $t\in T'$ be a sub-rate sink. The GEM $B_{t}$ is a full-rank
$r\times h_{t}$ matrix, $r>h_{t}$. Therefore, for any set of $h_{t}$
length-$r$ vectors spanning the same subspace as do the columns of
$B_{t}$, there exists a $h_{t}\times h_{t}$ matrix $D$ such that
the columns of $B_{t}D$ are these vectors. Formally, if $span\{v^{1},...,v^{h_{t}}\}=span\{B_{t}\}$,
then an $h_{t}\times h_{t}$ matrix $D$ necessarily exists such that
$B_{t}D=(v^{1}|v^{2}|...|v^{h_{t}})$. Since $U$ is an exact spanner
with respect to $\tilde{B}$, there exists a subset $U_{t}\subseteq U$
of size $h_{t}$ spanning the same subspace as the columns of $B_{t}$.
The choice $D_{t}=D$, and the matrix created by concatenating the
columns of $U$ as $\bar{B}$, we get that for every $t\in T'$, all
the columns of $B_{t}D_{t}$ are also columns in $\bar{B}$. The choice
of $R_{t}$ as the columns from $I_{m}$ pointing at the locations
of the columns of $B_{t}D_{t}$ in $\bar{B}$, we get that $B_{t}D_{t}=\bar{B}R_{t}$,
namely - $\bar{B}R_{t}$ is a BR-factorization of $B_{t}D_{t}$ for
every $t\in T'$. Thus, according to Lemma \ref{lem:LBR_fact_of_BD}-
any sub-rate sink $t\in$$T'$ can work at a decoding rate $h_{t}$.
\end{IEEEproof}
This corollary implies that when given a set of sub-rate sinks, in
order to enable all of them to work in their maximum possible decoding
rates (while not interfering with the full-rate, a.k.a eligible, sinks
to operate at rate $r$), it suffices to find a set $U$ of $r$ vectors
that: (a) are linearly independent; and (b) $U$ is an exact spanner
with respect to the GEMs of all the sub-rate sinks. Note that these
conditions do not imply that the cardinality of $U$ has to be $r$.
If indeed $|U|<r$, then $r-|U|$ linearly independent vectors may
be added to $U$ in order to create a linearly independent exact spanner
with cardinality $r$, as required for the precoding scheme described
in Corollary~\ref{cor:exact spanner_usage}.

\subsection{Three sub-rate sinks}

With the above understanding, the following lemma is fairly easy to
prove:
\begin{lem}
\label{lem:Given-three-sub-rate}Given three sub-rate sinks $t_{1},t_{2},t_{3}$,
with $h_{t_{i}}=2$ for $1\leq i\leq3$, in a linear multicast with
rate $r=3$, if $dim(B_{t_{1}}'\cap B_{t_{2}}'\cap B_{t_{3}}')=0$,
then there exists a linear multicast with rate $r=3$ allowing all
three sub-rate sinks to work at decoding rate $2$.
\end{lem}
This lemma is best understood by using geometric arguments. Thus,
given Corollary \ref{cor:exact spanner_usage}, for proving this lemma
it suffices to find a set $U$, of 3 linearly independent vectors,
such that every ``plane'' defined by $B_{t_{i}}$, i.e. $B_{t_{i}}'$
, is spanned by a combination of 2 of the vectors of $U$. Recalling
that every two planes intersect in a line - and given that all three
of planes intersect in a dot - we get that there are 3 different lines
of intersection - and choosing $U$ to be these 3 lines would satisfy
the condition of Corollary \ref{cor:exact spanner_usage}.
\begin{IEEEproof}
First, note that $B_{t_{i}}'$, $1\leq i\leq3$, represent 3 different
subspaces (otherwise, $dim(B_{t_{1}}'\cap B_{t_{2}}'\cap B_{t_{3}}')\geq1$,
since it is actually an intersection of only two planes). This implies
that for every $1\leq\{i,j\}\leq3,\:i\neq j$, one necessarily has
$dim(B_{t_{i}}'\cap B_{t_{j}}')=1;$ denote by $v_{ij}$ a vector
in $B_{t_{i}}'\cap B_{t_{j}}'$. Since $dim(B_{t_{1}}'\cap B_{t_{2}}'\cap B_{t_{3}}')=0$,
the set $U=\{v_{12},v_{13},v_{23}\}$ is a set of linearly independent
vectors, and since every vector is in the intersection of two subspaces,
$U$ is an exact spanner with respect to $B_{t_{i}},1\leq i\leq3$.
Therefore, according to Corollary \ref{cor:exact spanner_usage},
all 3 sub-rate sinks can work at a decoding rate of their max-flow
- $2$.
\end{IEEEproof}
Other cases pertaining to three sub-rate sinks have to be treated
carefully, as a set $U$ with the required characteristics does not
necessarily exist. The following example is aimed at demonstrating
that.
\begin{example}
\label{exa:(non-existence)}(non-existence) Let $B$ denote a linear
multicast of rate $r=3$, and let $T'=\{t_{1},t_{2},t_{3}\}$ be a
set of sub-rate sinks with $h_{t}=2$ for any $t\in T'$, with $B_{t_{1}}=\left(\begin{array}{cc}
1 & 0\\
0 & 0\\
0 & 1
\end{array}\right)$,$\:B_{t_{2}}=\left(\begin{array}{cc}
0 & 0\\
1 & 0\\
0 & 1
\end{array}\right)$,$\:B_{t_{3}}=\left(\begin{array}{cc}
1 & 0\\
1 & 0\\
0 & 1
\end{array}\right)$. In this case, $B_{t_{1}}'\cap B_{t_{2}}'\cap B_{t_{3}}'=span\left\{ v=\left(\begin{array}{c}
0\\
0\\
1
\end{array}\right)\right\} $, namely $dim(B_{t_{1}}'\cap B_{t_{1}}'\cap B_{t_{1}}')=1$. Unfortunately,
there is no possible choice of 3 vectors satisfying the conditions
for $U$ in Corollary \ref{cor:exact spanner_usage}. In fact, to
have the subspace $B_{t}'$ spanned by exactly two vectors of a set
$U'$ for every $t\in T'$, the set $U'$ has to consist of at least
$4$ vectors, making them necessarily linearly dependent.
\end{example}
In order to fully understand the difference between the cases in lemma
\ref{lem:Given-three-sub-rate} and example \ref{exa:(non-existence)},
examine the cases presented in figures \ref{fig:3-planes-intersecting}
and \ref{fig:3-planes-intersecting-dot}. A case similar to the one
of example \ref{exa:(non-existence)} is presented in figure \ref{fig:3-planes-intersecting}.
In this case, since the 3 planes intersect on a line, there is no
possible choice of 3 vectors which is an exact spanner. An exact spanner
for this example could be of the vectors $\{v,u_{1},u_{2},u_{3}\}$
marked in the figure. The case shown in figure \ref{fig:3-planes-intersecting-dot},
on the contrary, in which the three planes intersect in a dot (namely,
$B_{t_{1}}'\cap B_{t_{2}}'\cap B_{t_{3}}'$) - is a case in which
a set of 3 linearly independent vectors which is an exact spanner
exists, and it is the set of three vectors that are the intersection
of each 2 planes. In this case, the vectors marked as $\{u_{1},u_{2},u_{3}\}$
could be a legitimate choice of such an exact spanner.

\begin{figure}
\begin{minipage}[t]{0.45\columnwidth}%
\caption{\label{fig:3-planes-intersecting}3 planes intersecting on a line}
\begin{tikzpicture} 
\draw [->] (0,0) -- (1.5,0,0) node [right] {$x$}; 
\draw [->] (0,0) -- (0,1.5,0) node [above] {$y$}; 
\draw [->] (0,0) -- (0,0,1.5) node [below left] {$z$};  
\filldraw[fill=green!20!white, draw=green!50!black] (0,0) -- (1,0,0) -- (0,0,1) -- (0,0); 
\filldraw[fill=blue!20!white, draw=blue!50!black] (0,0) -- (-1,1,0) -- (0,0,0.5) -- (0,0); 
\draw [->] (0,0) -- (-0.5,.50,0) [yellow!90!black] node [left] [black] {$u_3$};  
\filldraw[fill=red!20!white, draw=red!50!black] (0,0) -- (0,1,0) -- (0,0,1) -- (0,0); 
\draw [->] (0,0) -- (0,0,0.5) [yellow!90!black] node [below] [black] {$v$}; 
\draw [->] (0,0) -- (0.5,0,0) [yellow!90!black] node [below] [black] {$u_1$}; 
\draw [->] (0,0) -- (0,0.5,0) [yellow!90!black] node [right] [black] {$u_2$}; 
\end{tikzpicture} %
\end{minipage}\hfill{}%
\begin{minipage}[t]{0.45\columnwidth}%
\caption{\label{fig:3-planes-intersecting-dot}3 planes intersecting in a dot}
\begin{tikzpicture} 
\draw [->] (0,0) -- (1.5,0,0) node [right] {$x$}; 
\draw [->] (0,0) -- (0,1.5,0) node [above] {$y$}; 
\draw [->] (0,0) -- (0,0,1.5) node [below left] {$z$};  
\filldraw[fill=green!20!white, draw=green!50!black] (0,0) -- (1,0,0) -- (0,0,1) -- (0,0); 
\filldraw[fill=blue!20!white, draw=blue!50!black] (0,0) -- (1,0,0) -- (0,1,0) -- (0,0); 
\filldraw[fill=red!20!white, draw=red!50!black] (0,0) -- (0,1,0) -- (0,0,1) -- (0,0); 
\draw [->] (0,0) -- (0.5,0,0) [yellow!90!black] node [below] [black] {$u_1$}; 
\draw [->] (0,0) -- (0,0.5,0) [yellow!90!black] node [left] [black] {$u_2$}; 
\draw [->] (0,0) -- (0,0,0.5) [yellow!90!black] node [left] [black] {$u_3$};  

\end{tikzpicture} %
\end{minipage}
\end{figure}

While the case of example \ref{exa:(non-existence)} will be dealt
with in more details in section \ref{sec:Partial-Sub-Rate}, where
partial sub-rate decoding is discussed, it emphasizes the need for
explicit conditions stating whether a set of sub-rate sinks can achieve
their max-flow simultaneously. The next section is concerned with
this matter.

\section{\label{sec:Full-Sub-Rate-coding}Full Sub-Rate coding - Definitions
and Construction Methodology}

In this section, results and insights gained while analyzing the case
$r=3$ are generalized to linear multicast operation at some arbitrary
rate $r=n$. The basic framework of sub-rate coding will be kept:
for a subset $T'\subset T$ of sinks, $t\in T'$, with $h_{t}<r$,
the goal is to allow each of these sinks to decode at a rate as close
to it's max-flow as possible, while any eligible sink $t$ in $T\smallsetminus T'$
can still decode the incoming information at rate $h_{t}=r$.

To this end, a sub-rate sink $t$ is declared as\emph{ fully sub-rate
decodable }under linear multicast $PB_{t}$, if it can decode incoming
information at rate $h_{t}$. In turn, this translates to a requirement
for the existence of matrices $D_{t}\in M_{h_{t}\times h_{t}}(F)$
and $R_{t}\in M_{n\times h_{t}}(F)$ -consisting of $h_{t}$ distinct
columns taken from $I_{n}$), such that $PB_{t}D_{t}=R_{t}$.

\subsection{Conditions for realizing full sub-rate coding}

In this section, we provide an algorithm for realizing the coding
scheme. To facilitate the derivation of general conditions sufficient
for guarantying that all sub-rate sinks in the set $T'$ are fully
sub-rate decodable, a few definitions will be given. Herein, all vectors
are of length $r$ over the field $F$, and all matrices are full
(column) rank matrices with $r$ rows and $r$ or less columns, over
the field $F$. Also note that when referring to a set of matrices
$\tilde{B}=\{B_{1},...,B_{k}\}$, in case two matrices $B_{i},B_{j}\in\tilde{B}$
span the same vector space, one matrix, say $B_{i}$, can be omitted
from $\tilde{B}$. Clearly, the corresponding sink $i$ may still
be fully sub-rate decodable, if so is the case with sink $j$ whose
matrix is $B_{j}$. Thus, w.l.o.g., assume that the matrices in the
set $\tilde{B}=\{B_{1},...,B_{k}\}$ do span $k$ different subspaces.
\begin{defn}
The \emph{commonality degree} of a vector $v$ \footnote{for that matter, we shall refer to a \emph{\small{}line} as a \emph{\small{}vector}
lying on that \emph{\small{}line}.} with respect to the set of matrices $\tilde{B}=\{B_{1},...,B_{k}\}$,
denoted by $comd_{\tilde{B}}(v)$, is the number of matrices $B_{i}$
in $\tilde{B}$ with $v\in B_{i}'$, the subspace spanned by $B_{i}$.
\begin{defn}
The \emph{commonality degree} of an arbitrary set of vectors $V=\{v_{1},...,v_{d}\}$
with respect to the set of matrices $\tilde{B}=\{B_{1},...,B_{k}\}$,
denoted by $comd_{\tilde{B}}(V)$, is the sum of the commonality degrees
of all the vectors in $V$ with respect to $\tilde{B}$, namely $comd_{\tilde{B}}(V)=\stackrel[i=1]{d}{\sum}comd_{\tilde{B}}(v_{i})$.
\begin{defn}
The \emph{commonality set size} of a set of matrices $\tilde{B}=\{B_{1},...,B_{k}\}$,
denoted by $comss(\tilde{B})$, is the minimum cardinality of a set
of vectors $V$, where $V$ is an exact spanner of $\tilde{B}$. Stated
formally: $comss(\tilde{B})=min\{|V|\ |\ V\ is\ an\ exact\ spanner\ of\ \tilde{B}\}$.
\end{defn}
\end{defn}
Equipped with the above definitions, the general condition guarantying
that all sub-rate sinks in a set $T'$ are fully sub-rate decodable
can be stated as follows:
\end{defn}
\begin{lem}
\label{lem:comss_dim_equality}Given a set of $k$ sub-rate sinks
$T'$ in a linear multicast $B$, consisting of a set of global encoding
matrices $\tilde{B}=\{B_{t_{1}},...,B_{t_{k}}\},$ if $comss(\tilde{B})=dim(\stackrel[i=1]{k}{\sum}B_{t_{i}}')$,
then all the sub-rate sinks in $T'$ are fully sub-rate decodable.
\end{lem}
\begin{IEEEproof}
By the definition of the commonality set size of $\tilde{B}$, there
exists an exact spanner $V$ with cardinality $|V|=comss(\tilde{B})$.
Showing that the vectors of $V$ are linearly independent will have
$V$ fulfill both conditions of Corollary \ref{cor:exact spanner_usage},
and hence conclude the proof. Since $V$ is an exact spanner of $\tilde{B}$,
it is clearly also a spanner of $\stackrel[i=1]{k}{\sum}B_{t_{i}}'$,
meaning that $dim(span\{V\})\geq dim(\stackrel[i=1]{k}{\sum}B_{t_{i}}')=|V|$,
where the equality follows from the condition of the lemma. Yet taking
into account that any set of vectors satisfies $|V|\geq dim(span\{V\})$,
necessarily results with $dim(span\{V\})=|V|$, meaning that all the
vectors of $V$ are linearly independent, and by corollary \ref{cor:exact spanner_usage},
all the sub-rate sinks in $T'$ are fully sub-rate decodable.
\end{IEEEproof}
Notably, finding the value\textbf{\uwave{ }}of $comss(\tilde{B})$
when given an arbitrary set of sub-rate sinks $T'$ - in order to
determine whether $P,\{D_{t}\}_{t\in T'}$ exist such that all the
sinks in $T'$ are fully sub-rate decodable - may not be a trivial
task.

Referring to \ref{lem:comss_dim_equality}, and taking in account
that it is clear that $comss(\tilde{B})\geq dim(\stackrel[i=1]{k}{\sum}B_{t_{i}}')$
(see e.g. Example \ref{exa:(non-existence)}), we shall provide sufficient
conditions under which a set of matrices $\tilde{B}$ satisfies $comss(\tilde{B})\leq dim(\stackrel[i=1]{k}{\sum}B_{t_{i}}')$.
Thus, in the following, we offer a (constructive) method for upper
bounding $comss(\tilde{B})$.
\begin{defn}
\label{def:The--commonality-set}The \emph{$c$-commonality set size}
of a set of matrices $\tilde{B}=\{B_{1},...,B_{k}\}$, denoted by
$comss_{\tilde{B}}(c)$ is defined as follows:
\begin{enumerate}
\item for $c=k$, $comss_{\tilde{B}}(k)=dim(\stackrel[i=1]{k}{\cap B_{i}'})$
\item for $c=k-1$, \\
$comss_{\tilde{B}}(k-1)=\stackrel[j=1]{k}{\sum}\left[dim(\underset{i\neq j}{\stackrel[i=1]{k}{\cap}B_{i}'})-dim(\stackrel[i=1]{k}{\cap B_{i}'})\right]$
\item for $c<k$, denote $d=(k-c)$, so: 
\newpage
\begin{strip}
\hrule\vspace{5mm}
$comss_{\tilde{B}}(c)=\stackrel[i_{1},...,i_{d}=1]{k}{\sum}\left[dim(\underset{i\neq i_{1},...,i_{d}}{\stackrel[i=1]{k}{\cap}B_{i}'})-\stackrel[j=i_{1}]{i_{d}}{\sum}dim(\underset{i\notin\{i_{1},...,i_{d}\}\backslash j}{\stackrel[i=1]{k}{\cap}B_{i}'})+\underset{\{j_{1},j_{2}\}\subseteq\{i_{1},...,i_{d}\}}{\sum}dim(\underset{i\notin\{i_{1},...,i_{d}\}\backslash\{j_{1},j_{2}\}}{\stackrel[i=1]{k}{\cap}B_{i}'})-...+(-1)^{d}\cdot dim(\stackrel[i=1]{k}{\cap B_{i}}')\right]$
\vspace{5mm}\hrule
\end{strip}
\end{enumerate}
Note that the dimensions counted by $comss_{\tilde{B}}(i),\ i>c$
are not re-counted in $comss_{\tilde{B}}(c)$. Note that for any $1\leq c\leq k$,
$comss_{\tilde{B}}(c)$ is well defined and is only based on the intersections
of relevant vector spaces resulting from $\tilde{B}$.
\end{defn}
\begin{claim}
Every element of the (external) sum in $comss_{\tilde{B}}(c)$ (of
either Def. \ref{def:The--commonality-set}.2 or \ref{def:The--commonality-set}.3)
is \uwave{indicative} of the dimension of a $c$-set intersection
(\uwave{where }\emph{\uwave{``\mbox{$c$}-set intersection''}}\uwave{
refers to an intersection of exactly }\emph{\uwave{\mbox{$c$}}}\uwave{
vector spaces}) perpendicular to any $(c+1)$-set intersection that
is a sub-vector space\textbf{\uwave{ }}\uwave{of the former}.
Therefore, given a set of vectors $V$ that span all the $(c+1)$-set
intersections, $comss_{\tilde{B}}(c)$ \uwave{provides} an upper
bound on the minimum number of vectors one has to add to $V$ so as
to span also all the $c$-set intersections. (Note that $comss_{\tilde{B}}(c)$
is not necessarily the minimum possible number, since there is no
requirement that all these additional vectors be linearly independent
among themselves).
\end{claim}
\begin{IEEEproof}
Straight forward from the definition and previous discussion.
\end{IEEEproof}
\begin{defn}
The \emph{commonality polynomial} of a set of matrices $\tilde{B}=\{B_{1},...,B_{k}\}$,
denoted $compol_{\tilde{B}}:\mathbb{N}^{k}\to\mathbb{N}$, is defined
as $compol_{\tilde{B}}(i_{1},...,i_{k})=\stackrel[c=1]{k}{\sum}c\cdot i_{c}$,
where $i_{c}=|\{v\in V\ |\ comd_{\tilde{B}}(v)=c\}|$, and $V$ is
an arbitrary set of vectors. $compol_{\tilde{B}}$ can be thought
of as receiving an arbitrary set of vectors $V$, and for every $1\leq c\leq k$,
calculating the number of vectors of commonality degree $c$ in the
set, substituting the result as $i_{c}$. 
\end{defn}
Defined as such, the outcome of $compol_{\tilde{B}}(i_{1},...,i_{k})$
would be the commonality degree of $V$ with respect to $\tilde{B}$
- namely $compol_{\tilde{B}}(i_{1},...,i_{k})=comd_{\tilde{B}}(V)$.

Note that in order for a set of vectors to be an exact spanner, all
entries of $compol_{\tilde{B}}$, $i_{c}$, - must be no greater than
the $c$-commonality set size of $\tilde{B}$; hence, when given a
set of vectors $V$, one is sure to have $compol_{\tilde{B}}:\{\{0,...,comss_{\tilde{B}}(1)\}\times...\times\{0,...,comss_{\tilde{B}}(k)\}\}\to\mathbb{N}$
as long as $V$ is an exact spanner, as shown in the following lemma.
\begin{lem}
\label{lem:compol_upper_bound}Given a set of matrices $\tilde{B}=\{B_{1},...,B_{k}\}$,
any vector $\bar{i}=(i_{1},...,i_{k})$, $\bar{i}\in\{\{0,...,comss_{\tilde{B}}(1)\}\times...\times\{0,...,comss_{\tilde{B}}(k)\}\}$,
that satisfies $compol_{\tilde{B}}(\bar{i})\geq\stackrel[c=1]{k}{\sum}dim(B_{c}')$,
means that there exists a set of vectors $V$ that constitute an exact
spanner of $\tilde{B}$, with $|\{v\in V\ |\ comd_{\tilde{B}}(v)=c\}|=i_{c}$
for every $1\leq c\leq k$. This means that $|V|=\stackrel[c=1]{k}{\sum}i_{c}$.
\end{lem}
\begin{IEEEproof}
For simplicity, a specific special case of the lemma will be proven,
of when $\bar{i}=(comss_{\tilde{B}}(1),...,comss_{\tilde{B}}(k))$.
This case is not the most general case, but its proof is simpler and
it is still covering many of the important cases .

First, notice that any subspace $B_{t}'$ can be presented as a sum
of subspaces, each created by the sum of intersections of $B_{t}'$
with a different number of other subspaces from $\tilde{B}$.

Let $B_{t}^{c}$ be the subspace that is a sum of all possible $c$-set
intersections that include $B_{t}'$. Clearly, $B_{t}'=B_{t}^{1}\supseteq B_{t}^{2}\supseteq...\supseteq B_{t}^{k}$.

Denote by $comdim_{\tilde{B}}(B_{t}',c)$ the $c$-commonality dimension
of $B_{t}'$ with respect to $\tilde{B}$, defined as $comdim_{\tilde{B}}(B_{t}',c)=dim(B_{t}^{c})-dim(B_{t}^{c+1});\ comdim_{\tilde{B}}(B_{t}',k)=dim(B_{t}^{k})$.
For each $1\leq c\leq k$, $comdim_{\tilde{B}}(B_{t}',c)$ represents
the contribution of the $c$-set intersections (that include $B_{t}'$)
to the dimension of $B_{t}'$ that had not been added with $(c+1)$-set
intersections. By definition, it is clear that $\stackrel[c=1]{k}{\sum}comdim_{\tilde{B}}(B_{t}',c)=dim(B_{t}')$.

Note that every $c$-set intersection (with $B_{t}'$ included) is
contributing its ``unique'' dimension to both $comdim_{\tilde{B}}(B_{t}',c)$
and $comss_{\tilde{B}}(c)$, with $comss_{\tilde{B}}(c)$ possibly
consisting additional $c$-set intersections (those that do not include
$B_{t}'$), leading to the fact that $comss_{\tilde{B}}(c)\geq comdim_{\tilde{B}}(B_{t}',c)$.
Since any $c$-set intersection repeats in $c$ sets, we get the relation
$c\cdot comss_{\tilde{B}}(c)\geq\stackrel[t=1]{k}{\sum}comdim_{\tilde{B}}(B_{t}',c)$,
which in turn results with $compol_{\tilde{B}}(\bar{i})=\stackrel[c=1]{k}{\sum}c\cdot comss_{\tilde{B}}(c)\geq\stackrel[t=1]{k}{\sum}\stackrel[c=1]{k}{\sum}comdim_{\tilde{B}}(B_{t}',c)=\stackrel[t=1]{k}{\sum}dim(B_{t}')$.
This means that $\bar{i}=(comss_{\tilde{B}}(1),...,comss_{\tilde{B}}(k))$
is indeed an adequate choice for this lemma.

Consequently, the construction of $V$ can be carried out according
to the guidelines below.

Begin with $V=\{\}$. The construction will be realized by running
through the commonality degrees $c$, beginning with $c=k$. For $c=k$,
choose $comss_{\tilde{B}}(k)=dim(\stackrel[i=1]{k}{\cap}B_{k}')$
linearly independent vectors from $\stackrel[i=1]{k}{\cap}B_{k}'$,
and add them to $V$.

For each $k-1\geq c\geq1$, choose $comss_{\tilde{B}}(c)$ vectors
by creating $c$-set intersections and choosing the number of vectors
required for spanning them according to the corresponding element
of the sum $comss_{\tilde{B}}(c)$ (in Definition \ref{def:The--commonality-set});
this process guarantees that vectors with higher commonality degree
required for spanning the intersection are already contained in $V$.
Note that by construction, there is necessarily a set of $comdim_{\tilde{B}}(B_{t}',c)$
vectors spanning $B_{t}^{c}$ for every $c$ (practically, any spanning
subset of $comdim_{\tilde{B}}(B_{t}',c)$ vectors from those chosen
by at least $c$-intersections including $B_{t}'$ will be adequate;
all of them are in $B_{t}'$ and there must be more than $comdim_{\tilde{B}}(B_{t}',c)$
of them). Overall, a total of $comss_{\tilde{B}}(c)$ vectors of commonality
degree of $c$ are systematically chosen in this step for each value
of $c$. Add them to $V$.

In conclusion, the process described above ensures that $|\{v\in V\ |\ comd_{\tilde{B}}(v)=c\}|=comss_{\tilde{B}}(c)=i_{c}$
for every $1\leq c\leq k$. Since for every $c$, one obtains a subset
of $\stackrel[\gamma=c]{k}{\sum}comdim_{\tilde{B}}(B_{t}',\gamma)$
vectors spanning $B_{t}^{c}$, one eventually has a set of size $dim(B_{t}')$
spanning $B_{t}'$, meaning that $V$ is an exact spanner - as required.
\end{IEEEproof}
\begin{thm}
\label{thm:The-Fully-Sub-Rate}The full sub-Rate Decodability Theorem:
Given a set of sub-rate sinks $T'$ in a linear multicast $B$, consisting
of a set of global encoding matrices $\tilde{B}=\{B_{t_{1}},...,B_{t_{k}}\}$
(every pair of matrices spanning two different subspaces), if there
exists a vector $\bar{i}=(i_{1},...,i_{k})$ , $\bar{i}\in\{\{0,...,comss_{\tilde{B}}(1)\}\times...\times\{0,...,comss_{\tilde{B}}(k)\}\}$,
so that $compol_{\tilde{B}}(\bar{i})\geq\stackrel[c=1]{k}{\sum}dim(B_{c}')$
and $\stackrel[c=1]{k}{\sum}i_{c}\leq dim(\stackrel[c=1]{k}{\sum}B_{c}')$
- then all the sub-rate sinks in $T'$ are fully sub-rate decodable.
\end{thm}
\begin{IEEEproof}
According to Lemma \ref{lem:compol_upper_bound}, since $compol_{\tilde{B}}(\bar{i})\geq\stackrel[c=1]{k}{\sum}dim(B_{c}')$,
there exists a set of vectors $V$ that is an exact spanner of $\tilde{B}$,
with $|V|=\stackrel[c=1]{k}{\sum}i_{c}$. Since $V$ is an exact spanner
of $\tilde{B}$, $comss(\tilde{B})\leq|V|=\stackrel[c=1]{k}{\sum}i_{c}\leq dim(\stackrel[c=1]{k}{\sum}B_{i}')$.
Also recall that $comss(\tilde{B})\geq dim(\stackrel[i=1]{k}{\sum}B_{t_{i}}')$,
and hence $comss(\tilde{B})=dim(\stackrel[i=1]{k}{\sum}B_{t_{i}}')$,
which according to Lemma \ref{lem:comss_dim_equality}, means that
all the sub-rate sinks in $T'$ are fully sub-rate decodable.
\end{IEEEproof}
To sum up this section, in a linear multicast network with rate $r$,
in order to determine whether all the sub-rate sinks in a set $T'$
with $\tilde{B}=\{B_{t_{1}},...,B_{t_{k}}\}$ are fully sub-rate decodable,
and find the proper linear multicast configuration fulfilling it,
we propose the following algorithm:
\begin{enumerate}
\item Find the proper domain for the polynomial $compol_{\tilde{B}}$: For
every $1\leq c\leq k$, compute $comss_{\tilde{B}}(c)$ according
to the definition. It is a closed form derived by all the possible
intersections of the subspaces $\{B_{t_{1}}',...,B_{t_{k}}'\}$.
\item Find a vector $\bar{i}\in\{\{0,...,comss_{\tilde{B}}(1)\}\times...\times\{0,...,comss_{\tilde{B}}(k)\}\}$
satisfying both conditions: $\stackrel[c=1]{k}{\sum}i_{c}\leq dim(\stackrel[c=1]{k}{\sum}B_{t_{c}}')$
and $compol_{\tilde{B}}(\bar{i})\geq\stackrel[c=1]{k}{\sum}dim(B_{t_{c}}')$.
If such a vector does not exist - this algorithm cannot guarantee
a proper linear multicast configuration 
\item Define an empty set $V=\{\}$.
\item For $c=k:-1:1$, proceed as follows:
\begin{enumerate}
\item Find $i_{c}$ linearly independent vectors with commonality degree
of $comss_{\tilde{B}}(c),$denoted by $V_{c}$, such that $V_{c}\cup V$
is a linearly independent set. 
\item Substitute $V\leftarrow V\cup V_{c}$.
\end{enumerate}
\item Since $|V|=\stackrel[c=1]{k}{\sum}i_{c}\leq dim(\stackrel[c=1]{k}{\sum}B_{i})$,
all the vectors in $V$ are linearly independent. If the number of
vectors in $V$ is smaller than $r$, find a set of $r-|V|$ vectors,
denoted $U$, such that $U+V$ span the $r$-dimensional vector space.
Accordingly, substitute $V\leftarrow U+V$.
\item Define $\bar{B}=mat(V)$, a matrix whose columns are the vectors of
$V$. $V$ is an exact spanner of $\tilde{B}$, allowing to find for
each $t\in T'$ an invertible $h_{t}\times h_{t}$ matrix $D_{t}$
satisfying $B_{t}D_{t}=\bar{B}R_{t}$, with $R_{t}$ being the columns
from $I_{r}$ pointing at the locations of the columns of $B_{t}D_{t}$
in $\bar{B}$ (see Definition \ref{cor:exact spanner_usage}). Since
$\bar{B}$ is invertible, it is possible to define the precoding matrix
$P=\bar{B}^{-1}$.
\item The result, a linear multicast $PB$, will allow any of the sub-rate
sinks $t\in T'$ to work at a decoding rate of its max-flow $h_{t}$
using the corresponding decoding matrix $D_{t}$.
\end{enumerate}

\subsection{Examples and Corollaries}

In order to provide the reader with some intuition regarding the use
of the the full sub-Rate Decodability theorem (Theorem \ref{thm:The-Fully-Sub-Rate})
and the algorithm that followed, this sub-section includes an example
for the usage of the algorithm, as well as two more general cases
that are immediate corollaries from the theorem. Appendix \ref{sec:Examples-for-the}
includes examples for more cases in which the FSRD theorem can be
easily applied in order to determine whether a set of sinks are fully
sub-rate decodable.
\begin{example}
Let $T'=\{t_{1},t_{2},t_{3}\}$ be three sub-rate sinks with $h_{t_{i}}=2$
for $1\leq i\leq3$ in a linear multicast $B$ with rate $r=3$ over
$GF(3)$. 
Let $\tilde{B}=\left\{ B_{t_{1}}=\left(\begin{array}{cc}
1 & 0\\
1 & 0\\
0 & 1
\end{array}\right),B_{t_{2}}=\left(\begin{array}{cc}
1 & 0\\
0 & 1\\
0 & 1
\end{array}\right),B_{t_{3}}=\left(\begin{array}{cc}
1 & 1\\
1 & 0\\
0 & 1
\end{array}\right)\right\} $ 
be the corresponding GEMs of these sub-rate sinks, then all the sub-rate
sinks in $T'$ are fully sub-rate decodable .
\end{example}
\begin{IEEEproof}
Since it is clear that $dim(B_{t_{1}}'\cap B_{t_{2}}'\cap B_{t_{3}}')=0$,
the example can be proven using Lemma \ref{lem:Given-three-sub-rate}.
In this section, the algorithm described at the end of the previous
section will be used for finding the modifications required in order
to find the proper linear multicast enabling full sub-rate decodability.
\begin{enumerate}
\item First, the domain for $compol_{\tilde{B}}$ is computed to be $\bar{i}\in\{\{0\}\times\{0,...3\}\times\{0\}\}$,
as follows:
\begin{enumerate}
\item First, for $c=3$, $comss_{\tilde{B}}(3)=dim(\stackrel[i=1]{3}{\cap}B_{i}')$.
It can be easily shown that $dim(\stackrel[i=1]{3}{\cap}B_{i}')=0$,
and so $comss_{\tilde{B}}(3)=0$. This means that the third entry
of $\bar{i}$ must be 0.
\item for $c=2$, \\
$comss_{\tilde{B}}(2)=\stackrel[j=1]{3}{\sum}\left[dim(\underset{i\neq j}{\stackrel[i=1]{3}{\cap}B_{i}'})-dim(\stackrel[i=1]{3}{\cap}B_{i}')\right]$ \\ $=\stackrel[j=1]{3}{\sum}dim(\underset{i\neq j}{\stackrel[i=1]{3}{\cap}B_{i}'})=dim(B_{t_{2}}'\cap B_{t_{3}}')+dim(B_{t_{1}}'\cap B_{t_{3}}')+dim(B_{t_{1}}'\cap B_{t_{2}}')$. \\
Since $B_{t_{2}}'\cap B_{t_{3}}'$=$\left(\begin{array}{c}
2\\
1\\
1
\end{array}\right)$, $B_{t_{1}}'\cap B_{t_{3}}'$=$\left(\begin{array}{c}
1\\
1\\
0
\end{array}\right)$ and $B_{t_{1}}'\cap B_{t_{2}}'$=$\left(\begin{array}{c}
1\\
1\\
1
\end{array}\right) $, $comss_{\tilde{B}}(2)=3,$ and so, the second entry of $\bar{i}$,
$i_{2}$ has to satisfy $0\leq i_{2}\leq3$.
\item for $c=1$, \\
$comss_{\tilde{B}}(1)=\stackrel[i_{1},i_{2}=1]{3}{\sum}$ \\ $\left[dim(\underset{i\neq i_{1},i_{2}}{\stackrel[i=1]{3}{\cap}}B_{i}')-\stackrel[j=i_{1}]{i_{2}}{\sum}dim(\underset{i\notin\{i_{1},i_{2}\}\backslash j}{\stackrel[i=1]{3}{\cap}}B_{i}')+dim(\stackrel[i=1]{3}{\cap}B_{i}')\right]$.
For each pair of indexes $\{i_{1},i_{2}\}$, denote the third index by $i_{3}$, then:  
%
\begin{strip}
\vspace{5mm}\hrule
\begin{center}
$comss_{\tilde{B}}(1)=\stackrel[i_{1},i_{2}=1]{3}{\sum}\left[dim(B_{i_{3}}')-dim(B_{i_{2}}'\cap B_{i_{3}}')-dim(B_{i_{1}}'\cap B_{i_{3}}')+dim(\stackrel[i=1]{3}{\cap}B_{i}')\right]$ \\ $=\stackrel[i=1]{3}{\sum}\left[dim(B_{t_{i}}')-\stackrel[j=1,j\neq i]{3}{\sum}dim(B_{t_{i}}'\cap B_{t_{j}}')\right]$.
\end{center}
\hrule \vspace{4mm}
\end{strip}
%

Since for every $1\leq i\leq3$, $dim(B_{t_{i}}')=2$, and for every
$1\leq j\leq3,\ j\neq i$, $dim(B_{t_{i}}'\cap B_{t_{j}}')=1$, we
get that $comss_{\tilde{B}}(1)=0.$ This means the the first entry
of $\bar{i}$ must also be 0.
\end{enumerate}
\item The vector $\bar{i}=(0,3,0)$ is the only vector in the domain that
satisfies both conditions of the FSRD theorem:
\begin{enumerate}
\item $\stackrel[c=1]{k}{\sum}B_{t_{c}}'$ is the 3-dimentional vector space,
hence $\stackrel[c=1]{3}{\sum}i_{c}=3\leqslant dim(\stackrel[c=1]{k}{\sum}B_{t_{c}}')$.
\item $compol_{\tilde{B}}(\bar{i})=2\cdot3=6\geqslant2+2+2=\stackrel[c=1]{k}{\sum}dim(B_{t_{c}}')$.
\end{enumerate}
\item Define $V=\{\}$.
\item Since $i_{3},i_{1}=0$, this stage only requires finding 3 linearly
independent vectors with $comss_{\tilde{B}}(2)$. Such vectors can
be easily found as the intersection vectors of each pair of sub-spaces.
This gives \\
$V_{2}=\left\{ \left(\begin{array}{c}
2\\
1\\
1
\end{array}\right),\left(\begin{array}{c}
1\\
1\\
0
\end{array}\right),\left(\begin{array}{c}
1\\
1\\
1
\end{array}\right)\right\} $, \\ 
and hence the output from this stage is \\
$V=\left\{ \left(\begin{array}{c}
2\\
1\\
1
\end{array}\right),\left(\begin{array}{c}
1\\
1\\
0
\end{array}\right),\left(\begin{array}{c}
1\\
1\\
1
\end{array}\right)\right\} $.
\item Since $|V|=3=r$, no vectors need to be added to $V$ in this stage.
\item Define $\bar{B}=mat(V)=\left(\begin{array}{ccc}
2 & 1 & 1\\
1 & 1 & 1\\
1 & 0 & 1
\end{array}\right)$.
\begin{enumerate}
\item For each $1\leq i\leq3$, we find matrices $R_{t_{i}},D_{t_{i}}$,
so that $B_{t_{i}}D_{t_{i}}=\bar{B}R_{t_{i}}$:
\begin{enumerate}
\item For $i=1$, $B_{t_{1}}=\left(\begin{array}{cc}
1 & 0\\
1 & 0\\
0 & 1
\end{array}\right)$, so $R_{t_{1}}=\left(\begin{array}{cc}
0 & 0\\
1 & 0\\
0 & 1
\end{array}\right)$, giving $\bar{B}R_{t_{1}}=\left(\begin{array}{cc}
1 & 1\\
1 & 1\\
0 & 1
\end{array}\right)$. The choice of $D_{t_{1}}=\left(\begin{array}{cc}
1 & 1\\
0 & 1
\end{array}\right)$, will provide $B_{t_{1}}D_{t_{1}}=\left(\begin{array}{cc}
1 & 1\\
1 & 1\\
0 & 1
\end{array}\right)$.
\item Similarly, for $i=2$, $B_{t_{2}}=\left(\begin{array}{cc}
1 & 0\\
0 & 1\\
0 & 1
\end{array}\right)$, and so $R_{t_{2}}=\left(\begin{array}{cc}
1 & 0\\
0 & 0\\
0 & 1
\end{array}\right)$ and $D_{t_{2}}=\left(\begin{array}{cc}
2 & 1\\
1 & 1
\end{array}\right)$ will provide $B_{t_{2}}D_{t_{2}}=\bar{B}R_{t_{2}}=\left(\begin{array}{cc}
2 & 1\\
1 & 1\\
1 & 1
\end{array}\right)$.
\item Finally, for $i=3$, $B_{t_{2}}=\left(\begin{array}{cc}
1 & 1\\
1 & 0\\
0 & 1
\end{array}\right)$, and so $R_{t_{3}}=\left(\begin{array}{cc}
1 & 0\\
0 & 1\\
0 & 0
\end{array}\right)$ and $D_{t_{3}}=\left(\begin{array}{cc}
1 & 1\\
1 & 0
\end{array}\right)$ will provide $B_{t_{i}}D_{t_{i}}=\bar{B}R_{t_{i}}=\left(\begin{array}{cc}
2 & 1\\
1 & 1\\
1 & 0
\end{array}\right)$.
\end{enumerate}
\item The precoding matrix is \\
$P=\bar{B}^{-1}=\left(\begin{array}{ccc}
2 & 1 & 1\\
1 & 1 & 1\\
1 & 0 & 1
\end{array}\right)^{-1}=\left(\begin{array}{ccc}
1 & 2 & 0\\
0 & 1 & 2\\
2 & 1 & 1
\end{array}\right)$.
\end{enumerate}
\item The result, $PB$, is the desired linear multicast.
\end{enumerate}
\end{IEEEproof}
\begin{cor}
Let $B$ denote a linear multicast of rate $r$. A single sub-rate
sink $T'=\{t\}$, with $h_{t}<r$ and $\tilde{B}=\{B_{t}\}$, is always
fully sub-rate decodable.
\end{cor}
\begin{IEEEproof}
While this case is not difficult to prove directly, the proof we provide
herein employs the FSRD theorem, Theorem \ref{thm:The-Fully-Sub-Rate}.
Since there is only one sub-rate sink $t$, we have $comss_{\tilde{B}}(1)=h_{t}$.
The choice of $\bar{i}$ is trivial - $\bar{i}=(h_{t})$. Then - $compol_{\tilde{B}}(\bar{i})=\stackrel[j\in1]{l}{\sum}j\cdot\bar{i}(l)=1\cdot h_{t}=dim(B_{t}')$,
and $\stackrel[j=1]{l}{\sum}\bar{i}(j)=h_{t}=dim(B_{t}')$. Therefore
- according to the FSRD theorem- $T'$ is fully sub-rate decodable.
\end{IEEEproof}
\begin{cor}
Let $B$ denote a linear multicast of rate $r$, and let $T'$ be
a set of $2\leq l\leq r$ sub-rate sinks with max-flow $h_{t}=r-1$,
consisting of a set of global encoding matrices $\tilde{B}=\{B_{1},...,B_{l}\}$
that span $l$ different subspaces. If $dim(\stackrel[t=1]{l}{\cap}B_{t}')=(r-l)$,
then $T'$ is fully sub-rate decodable.
\end{cor}
\begin{IEEEproof}
As above, it is not difficult to directly find a set of vectors fulfilling
the conditions of Corollary \ref{cor:exact spanner_usage} for this
case, still the proof below employs the FSRD theorem. Note that since
$dim(\stackrel[t=1]{l}{\cap}B_{t}')=(r-l)$, $commss_{\tilde{B}}(l)=(r-l)$.
Since every matrix in $\tilde{B}$ generates a unique ($r-1$ dimensional)
subspace, as $h_{t}=r-1$ , the fact that the intersection of all
the subspaces is of the smallest possible dimension implies that for
every $1\leq i\leq l$, $dim(\underset{t\neq i}{\stackrel[t=1]{l}{\cap}}B_{t}')=(r-l+1)$.
This assertion can be stated more mathematically as follows: since
every matrix in $\tilde{B}$ generates a unique $r-1$ dimensional
subspace, it is clear that for every $1\leq i\leq l$, $dim(\underset{t\neq i}{\stackrel[t=1]{l}{\cap}}B_{t}'+B_{i}')=r$
(since $dim(\stackrel[t=1]{l}{\sum}B_{t}')=r)$; then we have $dim(\underset{t\neq i}{\stackrel[t=1]{l}{\cap}}B_{t}')=dim(\underset{t\neq i}{\stackrel[t=1]{l}{\cap}}B_{t}'+B_{i}')-dim(B_{i}')+dim(\stackrel[t=1]{l}{\cap}B_{t}')=r-(r-1)+r-l=r-l+1$.
It follows from Definition \ref{def:The--commonality-set}.2 that
$commss_{\tilde{B}}(l-1)=l$. Consequently, the choice of $\bar{i}=(0,...0,l,r-l)$,
will produce $compol_{\tilde{B}}(\bar{i})=\stackrel[t=1]{l}{\sum}t\cdot i_{t}=(l-1)\cdot l+l\cdot(r-l)=l\cdot(r-1)=\stackrel[t=1]{l}{\sum}dim(B_{t}')$,
and $\stackrel[t=1]{l}{\sum}i_{t}=r=dim(\stackrel[t=1]{l}{\sum}B_{t}').$
Therefore - according to the FSRD theorem - $T'$ is fully sub-rate
decodable.
\end{IEEEproof}

\section{Partial Sub Rate Decoding Methodology\label{sec:Partial-Sub-Rate}}

\subsection{Motivation}

\subsubsection{Partial sub-rate decoding in $r=3$ for 3 sub-rate sinks}

We refer again to Example \ref{exa:(non-existence)}: let $B$ denote
a linear multicast of rate $r=3$, and let $T'=\{t_{1},t_{2},t_{3}\}$
be a set of sub-rate sinks with $h_{t}=2$ for every $t_{i}\in T'$.
Denote the corresponding matrices by $\tilde{B}=\{B_{1},B_{2},B_{3}\}$.
We have seen that when $dim(B_{1}'\cap B_{2}'\cap B_{3}')=1$, there
is no possible choice of 3 vectors that constitute an exact spanner
(Corollary \ref{cor:exact spanner_usage}). Indeed, in this case the
conditions for the full sub-rate decodability theorem are not satisfied,
since $comss_{\tilde{B}}(3)=1$ and $comss_{\tilde{B}}(2)=0$ (so
an exact spanner will require 4 vectors, out of which 3 vectors $v_{i}$
with $comd_{\tilde{B}}(v_{i})=1$).

Notwithstanding the foregoing, we next present an approach through
which it is possible to sub-rate decode all $t_{i}\in T'$, if one
is willing to compromise on the decoding rate. The following definition
precedes the main lemma of this subsection.
\begin{defn}
Define a $k$-degree BR-factorization of a $m\times n$ matrix $A$
$(m\geq n\geq k)$ as a multiplication of two matrices, $A=BR,$ such
that $B$ is a $m\times m$ full rank matrix and $R$ is an $m\times n$
matrix with $k$ (out of $n$) columns taken from $I_{m}$.
\end{defn}
\begin{lem}
\label{lem:Partial_SubRate3} Let $B$ denote a linear multicast of
rate $r=3$, and let $T'=\{t_{1},t_{2},t_{3}\}$ be a set of sub-rate
sinks with $h_{t}=2$ for every $t_{i}\in T'$. Denote the corresponding
GEM matrices by $\tilde{B}=\{B_{1},B_{2},B_{3}\}$. If \textup{$dim(B_{1}'\cap B_{2}'\cap B_{3}')=1$},
there exists a linear multicast of rate $r=3$ allowing every $t_{i}\in T'$
to work at a decoding rate of $r_{t_{i}}=5/3$, hence $t_{i}\in T'$
are partially sub-rate decodable.
\end{lem}
\begin{IEEEproof}
The idea behind this approach is to use a block of 3 network uses
(instead of a single use), which enables to find $5$-degree BR factorizations
with a single full rank matrix for all 3 of the GEM matrices of the
sinks in $T'$. With this factorization given, the idea implemented
by Lemma \ref{lem:LBR_fact_of_BD} is used in order to partially-decode
5 out of the 6 symbols each sink receives during the 3 network uses
- enabling them to work at decoding rate of $5/3$. The following
paragraph shows that mathematically.

Let $v_{4}\in B_{1}'\cap B_{2}'\cap B_{3}'$, and for every $1\leq i\leq3$,
$v_{i}$ is a vector so that $B_{t_{i}}'=span\{v_{i},v_{4}\}$. Note
that $V=\{v_{1},...,v_{4}\}$ is an exact spanner of $\tilde{B}$.
Let $\hat{B}$ denote the $3$-block linear multicast of $B$ (see
Subsection \ref{subsec:Sub-rate-Block-LNC}). Denote by $\hat{v_{i}}^{j}$
the vector obtained by substituting $v_{i}$ as the $j$-th block,
with all the other entries being $0$. For example, for $v_{3}=(v_{3,1},v_{3,2},v_{3,3})^{T}$,
we have $\hat{v_{3}}^{2}=\{0,\,0,\,0,\,v_{3,1},v_{3,2},v_{3,3},\,0,\,0,\,0\}^{T}$.
Note that the set $\hat{V}=\{\hat{v_{1}}^{1},\hat{v_{2}}^{1},\hat{v_{4}}^{1},\hat{v_{1}}^{2},\hat{v_{3}}^{2},\hat{v_{4}}^{2},\hat{v_{2}}^{3},\hat{v_{3}}^{3},\hat{v_{4}}^{3}\}$
is a linearly independent set of vectors, each of length $l\cdot r=3\cdot3=9$.
Defining $\hat{\bar{B}}$ as the matrix whose columns are the 9 vectors
from $\hat{V}$, clearly $\hat{\bar{B}}$ is invertible. Employing
$P=\hat{\bar{B}}^{-1}$as a precoding matrix will have each sub-rate
sink $t_{i}$ receiving $\hat{v}P\hat{B}_{t_{i}}$ at its input where
$\hat{v}$ denotes an input data vector $\hat{v}\in F^{l\cdot r}$.
Showing that $\hat{\bar{B}}R_{t_{i}}=$$\hat{B}_{t_{i}}D_{t_{i}}$
for some matrices $R_{t_{i}}$ ($l\cdot r\times l\cdot h_{t_{i}}=9\times6$
matrix with $5$ columns taken from $I_{9}$), and $D_{t_{i}}$ ($h_{t_{i}}\cdot l\times h_{t_{i}}\cdot l=6\times6$
matrix) will provide the appropriate $5$-degree BR factorizations.

Since for every $1\leq i\leq3$, $B_{t_{i}}'=span\{v_{i},v_{4}\}$,
the block form of $B_{t_{i}}'$ satisfies $\hat{B_{t_{i}}'}=span\{\hat{v_{i}}^{1},\hat{v_{4}}^{1},\hat{v_{i}}^{2},\hat{v_{4}}^{2},\hat{v_{i}}^{3},\hat{v_{4}}^{3}\}$.
For every $i$, let $D_{t_{i}}^{*}$ be a basis transformation matrix
such that $\hat{B}_{t_{i}}D_{t_{i}}^{*}=\left(\begin{array}{cccccc}
\hat{v_{i}}^{1} & \hat{v_{4}}^{1} & \hat{v_{i}}^{2} & \hat{v_{4}}^{2} & \hat{v_{i}}^{3} & \hat{v_{4}}^{3}\end{array}\right)$, hence, there are exactly $5$ columns of $\hat{\bar{B}}$ in $\hat{B}_{t_{i}}D_{t_{i}}^{*}$.
For every $i$, let $D_{t_{i}}$ be a matrix such that $\hat{B}_{t_{i}}D_{t_{i}}$
has these 5 columns from $\hat{\bar{B}}$, with the sixth column being
the zero vector. Then constructing $R_{t_{i}}$ so that it chooses
these 5 columns from $\hat{\bar{B}}$ and leaving the sixth column
as zero, provides the desired factorizaton $\hat{\bar{B}}R_{t_{i}}=$$\hat{B}_{t_{i}}D_{t_{i}}$.

With that understanding, by employing $D_{t_{i}}$ as the decoding
matrix for $t_{i}$ for every $i$: $\hat{v}P\hat{B_{t_{i}}}D_{t_{i}}=\hat{v}P\hat{\bar{B}}R_{t_{i}}=\hat{v}R_{t_{i}}$.
Namely, each sub-rate sink $t_{i}\in T'$ can decode $5$ symbols
every $3$ network uses, meaning that it is sub-rate decodable with
$r_{t_{i}}=5/l=5/3$.
\end{IEEEproof}
\begin{cor}
In a linear multicast of rate $r=3$, any set of up to 3 sub-rate
sinks with max-flow of $h_{t}=2$ can work at decoding rates of at
least $5/3$.
\end{cor}

\subsubsection{Partial sub-rate decoding in $r=3$ for 4 sub-rate sinks}

In view of the scenario discussed in the previous subsection one may
wonder whether it is also possible to use a similar approach for a
larger set $T'$ of sub-rate sinks. Herein, we answer this in the
affirmative by providing an example of achieving partial sub-rate
decoding in block linear multicasting for 4 sub-rate sinks.
\begin{lem}
\label{lem:Subratelem4}let $B$ denote a linear multicast of rate
r=3, and let $T'=\{t_{1},t_{2},t_{3},t_{4}\}$ be a set of sub-rate
sinks with $h_{t_{i}}=2$ for any $t_{i}\in T'$. Denote the corresponding
GEM matrices by $\tilde{B}=\{B_{1},B_{2},B_{3},B_{4}\}$. Assuming
that \textup{$dim(\stackrel[j=1]{3}{\cap}B_{t_{i_{j}}}')=0$} for
every $\{t_{i_{1}},t_{i_{2}},t_{i_{3}}\}$ there exists a linear multicast
of rate r=3 allowing every $t_{i}\in T'$ to work at a decoding rate
of $r_{t_{i}}=1.5$.
\end{lem}
\begin{IEEEproof}
The proof is quite similar to that of Lemma \ref{lem:Partial_SubRate3}.
Since the intersection of any set of 3 subspaces is of dimension 0,
it is not difficult to see that an exact spanning set is of minimum
size 4. Denote such spanning set by $V=\{v_{12},v_{13},v_{24},v_{34}\}$,
where $v_{ij}$ is a vector on the intersection line of the spaces
$B_{t_{i}}'$ and $B_{t_{j}}'$. Choosing a 4-block linear multicast
$\hat{B}$ with $\hat{V}=\{\hat{v_{12}}^{1},\hat{v_{13}}^{1},\hat{v_{23}}^{1},\hat{v_{13}}^{2},\hat{v_{24}}^{2},\hat{v_{34}}^{2},\hat{v_{12}}^{3},\hat{v_{34}}^{3},\hat{v_{24}}^{3},\hat{v_{12}}^{4},\hat{v_{13}}^{4},\hat{v_{34}}^{4}\}$,
would enable every sink to decode 6 symbols (the rest of this proof
follows the same line of arguments of Lemma \ref{lem:Partial_SubRate3}).
With 6 symbols decoded in 4 network uses, each sub-rate sink $t_{i}\in T'$
decoding rate is hence $r_{t_{i}}=6/4=1.5$.
\end{IEEEproof}

\subsection{Existence of partial sub-rate decoding in the general case }

From the previous lemmas, it seems that for any set of sub-rate sinks
$T'$, there is a linear multicast allowing them to work at some decoding
rate $h_{t}'\leq h_{t}$ . Next, this notion is formally substantiated.
\begin{defn}
Let $\tilde{B}=\{B_{1},...,B_{k}\}$ denote a set of $r$-row matrices
, each $B_{t}$ comprises of $h_{t}$ linearly independent columns.
A set of vectors $V=\{v_{1},...,v_{d}\}$ will be termed a \emph{partial
exact spanner} with respect to $\tilde{B}$, if for every $1\leq t\leq k$,
there is a subset of $V$ with $h_{t}'$ linearly independent vectors,
for some $1\leq h_{t}'\leq h_{t}$, denoted by $\{v^{1},...,v^{h_{t}'}\}$
with $span\{v^{1},...,v^{h_{t}'}\}\subseteq B_{t}'$.
\end{defn}
\begin{lem}
\label{lem:partial SubRate_general}Let $B\in M(F)_{r\times m}$ denote
a global encoding kernel for a linear multicast of rate r on the graph
$G=\{\mathcal{V},E\}$ with a source $s\in\mathcal{V}$ and set of
sinks $T\subseteq\mathcal{V}$. For every sink $t\in T,$ denote by
$B_{t}$ the GEM of $t$, $\tilde{B}=\{B_{1},...,B_{k}\}$, by $h_{t}$
its max-flow from the source $s$, and by $D_{t}$ its decoding matrix.
Let $T'\subset T$ be a set of sub-rate sinks, $\{t_{1},\ldots,t_{d}\}$.
If there exists a set of $r$ linearly independent vectors $U=\{v_{1},\text{\dots},v_{r}\}$
that constitutes a partial exact spanner with respect to $\tilde{B}$
with $\{h_{1}',...,h_{d}'\}$ spanning vectors (for each of the d
sinks), when for each $t\in T$, $1\leq h_{t}'\leq h_{t}$ - then
each sink $t$$\in T'$ can work at decoding rate of $h_{t}'$.
\end{lem}
\begin{IEEEproof}
let $t\in T'$ be a sub-rate sink. Its GEM $B_{t}$ is a full-rank
$r\times h_{t}$ matrix, $r>h_{t}$. Therefore, for a set $U_{t}$
of $h_{t}'$ vectors that span a subspace of $span\{columns\ of\ B_{t}\}$,
there exist a $h_{t}\times h_{t}'$ selection matrix $C_{t}$ (that
selects the $h_{t}'$ out of $h_{t}$ decodable input coordinates)
and $h_{t}'\times h_{t}'$ decoding matrix $D_{t}$, so that the columns
of $B_{t}C_{t}D_{t}$ are the vectors of $U_{t}$. Since for every
$t$, there is a $U_{t}\subset U$ that is such a set of $h_{t}$'
vectors, there also exist such matrices $D_{t}$ and $C_{t}$. By
the choice of $D_{t_{c}}=C_{t}D_{t}$, and the matrix created by concatenating
the columns of $U$ as$\bar{B}$, we get that for every $t\in T'$,
all the columns of $B_{t}D_{t}$ are columns of $\bar{B}$. The choice
of $R_{t}$ as the columns from $I_{r}$ pointing at the locations
of the columns of $B_{t}D_{t}$ in $\bar{B}$, yields $B_{t}D_{t}=\bar{B}R_{t}$,
namely - $\bar{B}R_{t}$ is a $h_{t}$-degree BR-factorization of
$B_{t}D_{t_{c}}$ for every $t\in T'$ . Therefore, each sub-rate
sink $t\in$$T'$ can work at decoding rate of $h_{t}'$.
\end{IEEEproof}
\begin{thm}
Let $B\in M(F)_{r\times m}$ be a global encoding kernel for a linear
multicast on the graph $G=\{\mathcal{V},E\}$ with the source $s\in\mathcal{V}$
and set of sinks $T\subseteq\mathcal{V}$. For every sink $t\in T,$
denote by $B_{t}$ the GEM of $t$, by $h_{t}$ its max-flow from
the source $s$, and by $D_{t}$ its decoding matrix. Let $T'\subset T$
be a set of sub-rate sinks, $\{t_{1},\ldots,t_{d}\}$. Each sub-rate
sink can work at some positive decoding rate.
\end{thm}
\begin{IEEEproof}
Denote by $\tilde{B}=\{B_{1},...,B_{k}\}$ a set of matrices derived
from the GEM of a corresponding set of sub-rate sinks; these matrices
span different subspaces. Let $V$ be an exact spanner of $\tilde{B}$
(choosing the exact spanner with minimal cardinality will provide
better results - in the sense of a smaller $l$, i.e. a shorter block
- but it is not mandatory). Let $dim(span\{V\})=d(V)$, and $l$ (the
block size) be the number of distinct subsets of $V$ that consist
$d(V)$ linearly independent vectors. Denote these subsets by $V_{i}\subset V$
for $1\leq i\leq l$. For each $V_{i}$, let $\bar{V_{i}}$ be a set
of $r$ linearly independent vectors (spanning the $r$-dimensional
vector space), with $V_{i}\subset\bar{V_{i}}$. Let $\hat{B}$ denote
the $l$-block linear multicast associated with $B$, and let $\hat{\tilde{B}}$
be the set of block matrices induced by $B$. Denote by $\hat{\ensuremath{\bar{V_{i}^{i}}}}$
the set of $r\cdot l$ column vectors created by substituting every
$v_{i}\in\bar{V_{i}}$ as the $i$-th block ($1\leq i\leq l$), with
all the other entries being $0$. Note that all the vectors in $\{\hat{\bar{V_{1}^{1}},...},\hat{\ensuremath{\bar{V_{l}^{l}\}}}}$are
linearly independent. Denote $\hat{V}=\stackrel[i=1]{l}{\bigcup}\hat{\ensuremath{\bar{V_{i}^{i}}}}$.
For every $t\in T'$, denote by $h_{t}'$ the number of columns in
$\hat{V}$ that are in the subspace spanned by $\hat{B}_{t}$ (Obviously,
$h_{t}\geq h_{t}'\geq1$, since there have been least $h_{t}$ vectors
spanning $B_{t}$ in $V$ - and the choice of all the distinct subsets
of $V$ that are linearly independent assures that they are included
at least once). Therefore, $\hat{V}$ is a partial exact spanner with
respect to $\hat{\tilde{B}}$, with $\{h_{1}',...,h_{d}'\}$. Finally,
employing Lemma \ref{lem:partial SubRate_general} concludes the proof.
\end{IEEEproof}

\subsubsection{Topics for Future Work}

The above theorem is arguing for the existence of a partial sub-rate
decoding methodology, but is leaving two important topics in that
respect for discussion:
\begin{enumerate}
\item \textbf{What is the minimum possible block length as a function of
the intersections between the sets in $\tilde{B}$?} Reducing the
block size $l$ is of interest, since it shortens the decoding delay,
requires smaller amount of memory at the end units and enables all
operations to be executed with smaller-size matrices. Note that one
may derive the minimum size of a ``valid'' exact spanner as the
minimum value of the sum $\stackrel[c=1]{k}{\sum}i_{c}$ when $\bar{i}=(i_{1},...,i_{k})$,
$\bar{i}\in\{\{0,...,comss_{\tilde{B}}(1)\}\times...\times\{0,...,comss_{\tilde{B}}(k)\}\}$,
that is satisfying $compol_{\tilde{B}}(\bar{i})\geq\stackrel[c=1]{k}{\sum}dim(B_{c}')$
(see Lemma \ref{lem:compol_upper_bound}). Albeit, it can in fact
be observed from the lemmas at the beginning of this section that
the size of the exact spanner is simply not sufficiently revealing,
even when taking into account the dimension of the sum space $dim(\underset{B_{i}\in\tilde{B}}{\sum B_{i})}$,
to determine the minimum possible block length; two different sets
of matrices, both with exact spanner of minimum size 4 and dimension
sum of 3 - had required two different block lengths - 3 and 4. Intuitively,
the block length should indeed be a function of the values of the
$c$-commonality set sizes of $\tilde{B}$ for all $1\leq c\leq|\tilde{B}|$,
yet an exact closed-form solution is not presented in this paper
\item \textbf{What are the maximum possible effective decoding rates in
which all the sub-rate sinks can operate simultaneously? }With $V$
being the minimal exact spanner, the intuition leads to the size $h_{t}'\sim\frac{h_{t}\cdot r}{|V|}$,
since in every network use the ``effective'' transmission rate to
the sub-rate sinks is $\frac{r}{|V|}$, and the max-flow to $t$ is
$h_{t}$. The reality is more complicated, as seen from the previous
lemmas \ref{lem:Partial_SubRate3} and \ref{lem:Subratelem4}, for
which the values of $r,h_{t},|V|$ are equal in both scenarios, yet
the decoding rates are different ($5/3$ and $1.5$, respectively).
The effective rate also depends on the values of the $c$-commonality
set sizes of $\tilde{B}$ for all $1\leq c\leq|\tilde{B}|$; finding
the exact dependency is left for future work.
\end{enumerate}

\section{Further Results and Relations to Other Methods}

All the results in this paper can be applied to vector network coding,
which means operation with blocks of size $m$ over a binary field
instead of the field of size $2^{m}$, referred to as scalar network
coding.

In this section, more involved combinations of sub-rate coding with
other multicast characteristics and methods are discussed.

\subsection{Sub-Rate Coding for Static Linear Network Codes}

Often, network configuration may change due to various reasons such
as node failures and communication-channel variations to name but
a few. A very important method in network coding relates to \emph{static
linear multicast}, aimed at facilitating the communication with all
the designated sinks under a few, different, network configurations
without having to change the network code when the network configuration
changes. This capability comes at the cost of increasing the field
size, hence reducing the communication rate, with the number of (predetermined)
configurations as a factor of the former increase.

A schematic algorithm for constructing static linear multicast was
presented in \cite{KoetterOct.2003,qifu2018}. First, for every sink
$t\in T$ of the network and every $\varepsilon\in\mathcal{\hat{\varepsilon}}$
configuration, an virtual sink $t^{\varepsilon}$ is created, with
$max-flow_{\varepsilon}(t)$ paths connecting it to the original source,
($max-flow_{\varepsilon}(t)$ denoting the max-flow from the source
to the real sink $t$ under the network configuration $\varepsilon$).
Then, the linear multicast is designed in the usual manner, with the
set of virtual sinks acting as the designated set of sinks, while
making sure that all the paths to each virtual sink $t^{\varepsilon}$
only employ links that are present in network configuration $\varepsilon$.
That approach yields $|\hat{\varepsilon}|$ $\varepsilon$-GEM matrices,
$\{B_{\varepsilon}\}_{\varepsilon\in\hat{\varepsilon}}$ - each representing
the linear multicast under one of the link-failure configurations.

There are two different approaches for combining sub-rate coding with
a static network code so as to provide static linear multicast:
\begin{enumerate}
\item When the different configurations are not a result of spontaneous
link failures, but rather expected changes in the network's topology,
it is reasonable to assume that the network source can keep track
of the instantaneous configuration in each network use. In this case,
for each configuration $\varepsilon$, a set of sub-rate sinks $T_{\varepsilon}'$,
each set accompanied by a precoding matrix $P_{\varepsilon}$, can
be defined with respect to the $\varepsilon$-GEM $B_{\varepsilon}$,
according to the methods presented in earlier sections of this work.
\item When the current network configuration is not monitored by the source,
the choice of the set of sub-rate sinks and precoding matrix cannot
change per configuration. In that case, a set of designated sub-rate
sinks $T'$ has to be fixed. For each sub-rate sink $t\in T'$, a
specific configuration $\varepsilon_{t}\in\hat{\varepsilon}$ has
to be chosen, and the sink's $\varepsilon$-GEMs, $B_{t,\varepsilon_{t}}$
is the matrix considered for all the computations presented in the
previous chapters. In this case, the precoding matrix $P$ is fixed,
and is computed as an invertible matrix using the set of matrices
$\{B_{t,\varepsilon_{t}}\}_{t\in T'}$. In general - each designated
sub-rate sink $t\in T'$ will only be able to apply the sub-rate decoding
scheme in the configuration $\varepsilon_{t}$ (or in other configurations
$\varepsilon\in\hat{\varepsilon}$ with $span\{B_{t,\varepsilon}\}=span\{B_{t,\varepsilon_{t}}\}$).
Note that - as in the non-static case described below - for every
designated real sink $t\in T$, for any configuration $\varepsilon$
where $max-flow_{\varepsilon}(t)\geq r$ ($r$ being the network's
operation rate), $B_{t,\varepsilon}$ is an invertible matrix, ensuring
that the precoding does not interfere with its decoding ability.
\end{enumerate}

\subsection{Sub-Rate Coding for Variable-Rate Linear Network Codes}

An interesting method to combine with sub-rate coding is the so-called
\emph{variable-rate network codin}g \cite{Fong2006.Oct.2006,qifu2018}.
In essence, this method enables to work with the same linear multicast
under different rates of symbol-generation from the source (limited
by the number of outgoing links from the source). For example, if
the source has $5$ outgoing links, then linear multicast using variable-rate
network coding will allow it to operate at any rate $r\leq5$, making
the transmission decodable by all sinks whose max-flow is at least
$r$. In contrary to the previous methods, this method does not necessitate
a larger field size, hence it entails no reduction in rate.

The construction method of variable rate linear multicast is carried
out much the same as linear multicast, only using a slightly different
input graph, leading to a change in both the path-finding stage and
the values calculation stage of the linear multicast. The algorithm
results with a Local Encoding Kernel, but note that the traditional
GEM $B$ can not be well defined, because the number of its rows equals
the symbols generation rate $r$, which is not constant. Alternatively,
for every network operation rate $1\leq q\leq r$, $B_{q}$, a $q$\emph{-GEM
}is defined, as a $q\times|E|$ matrix that represents the network
code operation under the rate $q$. In order to combine sub-rate coding
with variable-rate coding, it is necessary to refer to each rate separately.
Namely, for every operation rate $q$ a different set of sub-rate
sinks has to be defined with respect to $B_{q}$, and a corresponding
$q\times q$ precoding matrix $P_{q}$ has to be derived. Since the
network operation rate is determined by the network source, it is
only natural that it can also adjust the precoding scheme accordingly.

Note that since it is necessary to adjust the sub-rate coding scheme
for each rate, one may choose different sinks for each rate. This
makes perfect sense, as for each operation rate, different sinks have
max-flow from the source that is smaller than the network operation
rate; i.e., the set of potential sub-rate sinks changes with the network
operation rate.

\subsection{Sub-Rate Coding as Alternative to Sinks Addition}

Finally, we tackle a more intricate issue - which, in a way, generalizes
the above two cases. As more nodes in a network get designated as
sinks, the larger is the field-size over which calculations need to
be performed, hence - in many cases - resulting with reduction in
communication rate to all the sinks in the network. An interesting
approach to alleviate this problem, while still allowing communication
with many (designated) nodes without having to increase the field-size,
is to avoid regarding these designated nodes as standard sinks, but
rather as \emph{consequential sub-rate sinks}. To this end, we propose
to construct the linear multicast while considering some designated
sinks (not as standard sinks but rather) as ``intermediate'' nodes.
Then, using the effective rate allowed by the network code, each such
intermediate node will be approached as sub-rate sink and operate
at its \emph{consequential }max-flow, which is basically the number
of linearly independent inputs to this node as entailed by the linear
multicast.

In certain cases this approach may actually improve the communication
rate with the designated nodes compared to the rate achievable by
regarding them as (standard) sinks - especially when working with
a small set of sinks (hence a small field). Finally, since increased
field size always entails reduced communication rate to all the sinks,
the proposed approach may prove a viable modification to any design
of a network code.

The following claim offers a simple sanity-check for identifying whether
a \emph{single} node should be regarded as a standard sink, or as
a consequential sub-rate sink. Just recall that, as explained above,
this is not the only consideration in favor of employing the sub-rate
coding methodology.
\begin{claim}
\label{claim:conseq sub rate sinks}Let $B\in M(F)_{r\times m}$ be
a global encoding kernel for a linear multicast on the graph $G=\{\mathcal{V},E\}$
with the source $s\in\mathcal{V}$ and set of sinks $T\subseteq\mathcal{V}$.
For every \emph{node} $t\in\mathcal{V}$, denote by $h_{t}$ its max-flow
from the source $s$, and by $B_{t}$ the GEM of $t$, which is the
matrix whose columns, taken from $B$, represent the links entering
$t$, just as the GEM is defined for a ``standard'' sink. For a
node $t\notin T$ let $r_{t}$ denote the consequential max-flow,
namely the number of linearly independent columns in $B_{t}$ (for
standard sinks, $r_{t}=h_{t}$, but in general $r_{t}\leq h_{t}$).
Denote by $F'$ the field required for considering the set of sinks
$T^{*}=T\cup\{t\}$. If $r_{t}\cdot\frac{1}{\lceil log_{2}|F|\rceil}\geq h_{t}\frac{1}{\lceil log_{2}|F'|\rceil}$,
then considering $t$ as a sink would reduce its communication rate
compared to considering it as a consequential sub-rate sink.
\end{claim}
\begin{IEEEproof}
Consider two different cases:
\begin{enumerate}
\item Referring to $t$ as a consequential sub-rate sink, would allow it
to decode $r_{t}$ symbols with each network use. Since the network
messages under the network code represented by $B$ are elements of
the field $F$, each symbol is represented by $\lceil log_{2}|F|\rceil$
bits - meaning that the communication rate to $t$ as a consequential
sub-rate sink is $r_{t}\cdot\frac{1}{\lceil log_{2}|F|\rceil}$ bits
per network use.
\item Considering $t$ as a sink (i.e. - calculating a new linear multicast,
with $T^{*}$ being the set of sinks) will allow $t$ to decode at
most $h_{t}$ symbols per network use. In this case, the symbols are
elements of the field $F'$ (clearly $F'>F$), hence the communication
rate to $t$ will be at most $h_{t}\frac{1}{\lceil log_{2}|F'|\rceil}$
bits per network use. Notably, that the network's max-flow in this
case is $\underset{\tau\in T^{*}}{min}\{max-flow\ between\ s\ and\ \tau\}$,
and may very well be smaller than $h_{t}$.
\end{enumerate}
Thus, if $r_{t}\cdot\frac{1}{\lceil log_{2}|F|\rceil}\geq h_{t}\frac{1}{\lceil log_{2}|F'|\rceil}$,
then the number of decoded bits per network use is higher when referring
to $t$ as a consequential sub-rate sink.
\end{IEEEproof}
Note that, while using claim \ref{claim:conseq sub rate sinks} is
straight-forward for a single consequential sub-rate sink, using it
for more than one sink requires that the set of potential consequential
sub-rate sinks satisfy the conditions given in Section \ref{sec:Full-Sub-Rate-coding}
of this paper, namely that this set could be fully sub-rate decodable.

\begin{figure}[htb]
\begin{center}
\includegraphics[width=1\columnwidth, angle=0]{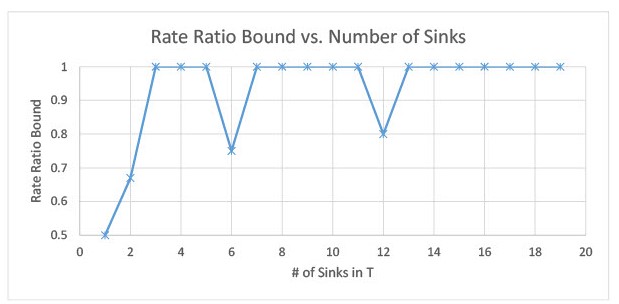}
\caption{Rate Ratio Bound}
\label{Fig:RateRatio}
\end{center}
\end{figure}

Figure \ref{Fig:RateRatio} shows the function $\frac{\lceil log_{2}|F|\rceil}{\lceil log_{2}|F'|\rceil}$,
termed the rate ratio bound, as a function of the number of sinks.
As the requirement for the existence of a linear multicast is that
$|F|>|T|$, for every $T$, the cardinality of the field to work with
is the smallest prime number greater than $T$. If the ratio $\frac{r_{t}}{h_{t}}$
is above the curve in the figure, then considering $t$ as a sink
will actually reduce its communication rate as opposed to considering
it as a sub-rate sink.

As seen in the figure, some values of the rate ratio require $\frac{r_{t}}{h_{t}}$
to be quite higher than others in order for sub-rate sink to be superior
to standard sink designation.

\section{Conclusions}

\emph{Sub-rate coding} is introduced and thoroughly studied. It amounts
to network (pre)-coding at the source side and decoding of partial
information by a set of sinks whose max-flow is smaller than the network
operation rate. This work is concerned with a single-source, finite
acyclic \emph{scalar} network, although it is straight-forward to
apply the results to vector networks.

Sub-rate coding methodology does not impact the transmission rate
towards sinks whose max-flow permits full-rate decoding. This makes
sub-rate coding a viable add-on to any network. Applications of sub-rate
coding include any scenario that employs multicasting to communicate
with different users experiencing different channel qualities, or
otherwise when they require different volumes of the data. It is especially
valuable when a small number of sinks have a wide dynamic range of
max-flows.

For achieving \emph{full sub-rate decodability}, a general and sufficient
condition allowing each sub-rate sink to operate at its optimal rate
(its max-flow from the source) is introduced. Furthermore, an algorithm
is proposed for determining if full sub-rate decodability is realizable
for a given set of sub-rate sinks; the necessary modifications to
the existing linear network code are detailed, bearing no impact on
the original network sinks.

With respect to \emph{partial sub-rate decodability}, it was proven
that for any set of sub-rate sinks there exists an effective sub-rate
coding scheme that is based on the transmission of a sequence of messages
thus resulting with a linear sub-rate network block-code. A few examples
are provided.

Finally, the combination of \emph{sub-rate coding} with other important
linear network codes is briefly discussed. Proper ways to combine
sub-rate coding with static linear network codes and with variable-rate
linear network codes are presented. It is also shown that in some
cases, sub-rate coding can be a preferable solution when additional
sinks are to added to an existing network.

\newpage
\appendices{}

\section{\label{sec:Examples-for-the}FSRD Theorem Examples}

To demonstrate the usage of the FSRD theorem (Theorem \ref{thm:The-Fully-Sub-Rate})
in some more detail, and provide better intuition as to the actual
meaning of each of the measures, many examples for possible types
of GEMs (given a set of sinks) are summarized in Table \ref{tab:GEMs-types-and}.
For each type of GEM set $\tilde{B}=\{B_{t_{1}},...,B_{t_{k}}\}$,
the values $comss_{\tilde{B}}(c)$ for $1\leq c\leq k$, $\stackrel[c=1]{k}{\sum}dim(B_{c}')$
and $dim(\stackrel[c=1]{k}{\sum}B_{c}')$ are calculated, and then
a vector $\bar{i}$ that satisfies the FSRD theorem conditions (namely,
$compol_{\tilde{B}}(\bar{i})\geq\stackrel[c=1]{k}{\sum}dim(B_{c}')$
and $\stackrel[c=1]{k}{\sum}i_{c}\leq dim(\stackrel[c=1]{k}{\sum}B_{c}')$)
is suggested, if such a vector exists.

\bibliographystyle{plain}
\bibliography{bibliography_SR_LNC}

\onecolumn
\begin{table}[tp]
\caption{\label{tab:GEMs-types-and}GEMs types and their corresponding $\bar{i}$
for FSRD theorem}
\vspace{-0.5cm}
\begin{longtable}{|>{\centering}p{3cm}|>{\centering}p{3cm}|>{\centering}p{3cm}|>{\centering}p{2cm}|>{\centering}p{2cm}|>{\centering}p{3cm}|}
\hline 
GEMs & GEMs intersections & $comss_{\tilde{B}}(c)$ for $1\leq c\leq k$ & $\stackrel[c=1]{k}{\sum}dim(B_{c}')$ & $dim(\stackrel[c=1]{k}{\sum}B_{c}')$ & Possible FSRD $\bar{i}$\tabularnewline
\hline 
\hline 
A vector $B_{1}=v_{1}$ with $r\geq2$ & $dim(B_{1}')=1$ & $comss_{\tilde{B}}(1)=1$ & $dim(B_{1}')=1$ & $dim(B_{1}')=1$ & $\bar{i}=(1)$\tabularnewline
\hline 
Two vectors $B_{1}=v_{1},\ B_{2}=v_{2}$ with $r\geq2$ & $dim(B_{1}'\cap B_{2}')=0$ & $comss_{\tilde{B}}(2)=0$, $comss_{\tilde{B}}(1)=2$ & $dim(B_{1}')+dim(B_{2}')=2$ & $dim(B_{1}'+B_{2}')=2$ & $\bar{i}=(2,0)$\tabularnewline
\hline 
Three vectors $B_{1}=v_{1},\ B_{2}=v_{2},\ B_{3}=v_{3}$ with $r\geq2$ & $dim(B_{i}'\cap B_{j}')=0$ but $\{v_{1},v_{2},v_{3}\}$ are linearly
dependent & $comss_{\tilde{B}}(3)=0$, $comss_{\tilde{B}}(2)=0$, $comss_{\tilde{B}}(1)=3$ & $3$ & $2$ & Impossible - the only eligible$\bar{i}$ with $compol_{\tilde{B}}(\bar{i})\geq3$
is $\bar{i}=(3,0,0)$, for which $\stackrel[c=1]{3}{\sum}i_{c}>2$\tabularnewline
\hline 
Two 2-column matrices $B_{1},B_{2}$ with $r\geq3$ & $dim(B_{1}'\cap B_{2}')=1$ & $comss_{\tilde{B}}(2)=1$, $comss_{\tilde{B}}(1)=2$ & $4$ & $3$ & $\bar{i}=(2,1)$\tabularnewline
\hline 
Two 2-column matrices $B_{1},B_{2}$ and a vector $B_{3}=v_{1}$with
$r\geq4$ & $dim(B_{1}'\cap B_{2}')=1$, $dim(B_{1}'\cap B_{3}')=dim(B_{2}'\cap B_{3}')=0$ & $comss_{\tilde{B}}(3)=0$, $comss_{\tilde{B}}(2)=1$, $comss_{\tilde{B}}(1)=3$ & $5$ & $4$ (Note that for $r=3$, $dim(\stackrel[c=1]{3}{\sum}B_{c}')=3$,
resulting in no FSRD possibility) & $\bar{i}=(3,1,0)$\tabularnewline
\hline 
Four 2-column matrices $B_{1},...,B_{4}$, with $r=3$ & For each $i\neq j$, $dim(B_{i}'\cap B_{j}')=1$, and for each $i\neq j\neq k$,
$dim(B_{i}'\cap B_{j}'\cap B_{k}')=0$ & $comss_{\tilde{B}}(4)=0$, $comss_{\tilde{B}}(3)=0$, $comss_{\tilde{B}}(2)=6$,
$comss_{\tilde{B}}(1)=0$ & 8 & 3 & Impossible - in order to have $compol_{\tilde{B}}(\bar{i})\geq8$,
necessarily $i_{2}\geq4$, giving $\stackrel[c=1]{4}{\sum}i_{c}>3$\tabularnewline
\hline 
Three 3-column matrices $B_{1},B_{2},B_{3}$ with $r\geq4$ & $dim(\stackrel[k=1]{3}{\cap}B_{k}')=1$, and for each $i\neq j$,
$dim(B_{i}\cap B_{j})=2$ & $comss_{\tilde{B}}(3)=1$, $comss_{\tilde{B}}(2)=3$, $comss_{\tilde{B}}(1)=0$ & $9$ & $4$ & $\bar{i}=(0,3,1)$\tabularnewline
\hline 
Three 3-column matrices $B_{1},B_{2},B_{3}$ with $r\geq4$ & $dim(\stackrel[k=1]{3}{\cap}B_{k}')=2$, and for each $i\neq j$,
$dim(B_{i}\cap B_{j})=2$ & $comss_{\tilde{B}}(3)=2$, $comss_{\tilde{B}}(2)=0$, $comss_{\tilde{B}}(1)=3$ & $9$ & $4$ (For $r>4$, may also be $5$) & Only possible if $dim(\stackrel[c=1]{k}{\sum}B_{c}')=5$, with $\bar{i=(3,0,2)}$\tabularnewline
\hline 
Four 3-column matrices $B_{1},...,B_{4}$, with $r\geq4$ & $dim(\stackrel[k=1]{4}{\cap}B_{k}')=0$, for each $i$ $dim(\stackrel[k=1,k\neq i]{4}{\cap}B_{k}')=1$,
and for every $i\neq j$ $dim(B_{i}\cap B_{j})=2$ & $comss_{\tilde{B}}(4)=0$, $comss_{\tilde{B}}(3)=4$, $comss_{\tilde{B}}(2)=0$,
$comss_{\tilde{B}}(1)=0$ & $12$ & $4$ & $\bar{i}=(0,0,4,0)$\tabularnewline
\hline 
Four 3-column matrices $B_{1},...,B_{4}$, with $r\geq4$ & $dim(\stackrel[k=1]{4}{\cap}B_{k}')=1$, for each $i$ $dim(\stackrel[k=1,k\neq i]{4}{\cap}B_{k}')=1$,
and for every $i\neq j$ $dim(B_{i}\cap B_{j})=2$ & $comss_{\tilde{B}}(4)=1$, $comss_{\tilde{B}}(3)=0$, $comss_{\tilde{B}}(2)=4$,
$comss_{\tilde{B}}(1)=0$ & 12 & 4 & Impossible - the only eligible$\bar{i}$ with $compol_{\tilde{B}}(\bar{i})\geq12$
is $\bar{i}=(0,4,0,1)$, for which $\stackrel[c=1]{3}{\sum}i_{c}>4$\tabularnewline
\hline 
$n$ matrices with $n-1$ columns, $B_{1},...,B_{n}$, with $r=n$ & $dim(\stackrel[k=1]{n}{\cap}B_{k}')=0$ & $comss_{\tilde{B}}(n)=0$, $comss_{\tilde{B}}(n-1)=n$, $comss_{\tilde{B}}(n-2)=...=comss_{\tilde{B}}(1)=0$ & $n\cdot(n-1)$ & $n$ & $\bar{i}=(0,...,0,n,0)$\tabularnewline
\hline 
\end{longtable}
\end{table}

\end{document}